\documentclass[a4paper,11pt]{article}

\usepackage{amsmath,amsfonts,amssymb,caption,graphicx}
\usepackage{ushort}
\usepackage{makeidx}
\usepackage{float}
\usepackage{accents}
\usepackage{color}
\usepackage{xcolor}
\usepackage{framed}
\usepackage{versions}
\usepackage{emptypage}
\usepackage{wrapfig}
\usepackage[T1]{fontenc}
\usepackage[utf8]{inputenc}
\usepackage{titlesec}
\usepackage{fancyhdr}
\usepackage{extramarks}
\usepackage[bookmarks]{hyperref}
\usepackage{hyperref}
\usepackage{bbm}
\usepackage{enumitem}
\usepackage[page]{appendix}

\usepackage{ragged2e}

\hypersetup{
    colorlinks=true,   	
    linkcolor=red,      
    citecolor = [rgb]{0 0.7 0},   	
    filecolor=magenta, 	
    urlcolor=blue
}

\def\transpose{{\hbox{\tiny\it T}}}
\def\transposed{\intercal}

\newcommand{\Xp}{\mbox{\boldmath $X$}}
\newcommand{\Up}{\mbox{\boldmath $U$}}

\newcommand{\bfone}{{\bf 1}}
\newcommand{\Xps}{\mbox{\scriptsize\boldmath $X$}}
\newcommand{\Ups}{\mbox{\scriptsize\boldmath $U$}}

\newcommand{\Yp}{\mbox{\boldmath $Y$}}

\newcommand{\subsqrt}{_{\sqrt{}}}

\newcommand{\RL}{{\mathbb R}}

\newcommand{\IND}{{\mathbb I}}
\newcommand{\BBP}{{\mathbb P}}

\newcommand{\VAR}{\mbox{\rm Var}}

\def\ba{\begin{align}}
\def\ea{\end{align}}
\def\ban{\begin{align*}}
\def\ean{\end{align*}}

\def\be{\begin{eqnarray}}
\def\ee{\end{eqnarray}}
\def\ben{\begin{eqnarray*}}
\def\een{\end{eqnarray*}}

\def\bqq{\begin{equation}}
\def\eqq{\end{equation}}
\def\bqqn{\begin{equation*}}
\def\eqqn{\end{equation*}}

\def\sq{$\Box$}

\def\qed{\ifmmode\sq\else{\unskip\nobreak\hfil
\penalty50\hskip1em\null\nobreak\hfil\sq
\parfillskip=0pt\finalhyphendemerits=0\endgraf}\fi\par\medbreak}

\newsavebox{\junk}
\savebox{\junk}[1.6mm]{\hbox{$|\!|\!|$}}

\def\limsup{\mathop{\rm lim\ sup}}
\def\liminf{\mathop{\rm lim\ inf}}

\def\bfP{{\bf P}}
\def\bfQ{{\bf Q}}
\def\bfR{{\bf R}}
\def\bfS{{\bf S}}

\def\til={{\widetilde =}}

\def\clC{{\cal C}}
\def\clD{{\cal D}}

\def\clH{{\cal H}}

\def\clP{{\cal P}}
\def\clQ{{\cal Q}}

\def\clS{{\cal S}}

 \def\eq#1/{(\ref{#1})}

\newtheorem{theorem}{Theorem}[section]

\newtheorem{proposition}[theorem]{Proposition}

\def\eq#1/{(\ref{e:#1})}

\def\bdes{\begin{description}}
\def\edes{\end{description}}

\def\notes#1{}

\definecolor{mag}{rgb}{0.7,0,0.3}
\definecolor{dgreen}{rgb}{0.1,0.5,0.1}
\definecolor{dred}{rgb}{.8,0,0}
\definecolor{gray}{rgb}{.8,.8,.8}
\definecolor{brown}{rgb}{0.6451,0.3706,0.1745}

\title{The Sample Complexity of Lossless Data Compression}

\author{
Terence Viaud
\and 
Ioannis Kontoyiannis
}

\date{\today}

\begin{document}

\maketitle

\footnotetext{The authors are with the
Statistical Laboratory, Centre for Mathematical Sciences, 
University of Cambridge, Wilberforce Road, Cambridge CB3 0WB, 
UK. Email: 
\texttt{tv329@cam.ac.uk},
\texttt{yiannis@maths.cam.ac.uk}.\\
This work was supported in part
by the EPSRC-funded INFORMED-AI project EP/Y028732/1.}

\begin{abstract}
A new framework is introduced for examining
and evaluating the fundamental limits of lossless
data compression, which emphasizes genuinely 
non-asymptotic results. The {\em sample
complexity} of compressing a given
source is defined as the
smallest blocklength at which it is possible
to compress that source 
at a specifically
constrained rate and to within a specified 
excess-rate probability.
This formulation parallels
corresponding developments in statistics
and computer science, and 
it facilitates the use of existing
results on the sample complexity of
various hypothesis testing problems.
For arbitrary sources, the sample complexity
of general variable-length compressors 
is shown to be tightly coupled with the
sample complexity of prefix-free codes
and fixed-length codes.
For memoryless sources,
it is shown that the sample complexity
is characterized not by the source
entropy, but by its 
R\'{e}nyi entropy of order~$1/2$.
Non-asymptotic bounds on the sample 
complexity are obtained, 
with explicit constants.
Generalizations to Markov sources are established,
showing that the sample complexity is determined
by the source's R\'{e}nyi entropy rate of
order~$1/2$.
Finally, bounds on the sample complexity of universal
data compression are developed for 
families of memoryless sources.
There, the sample complexity is characterized
by the minimum R\'{e}nyi divergence of order~$1/2$
between elements of the family 
and the uniform distribution.
The connection of this problem with identity
testing and with the associated separation rates
is explored and discussed.
\end{abstract}


\newpage

\section{Introduction}
\subsection{Lossless compressors}
\label{s:generalf}

A {\em variable-length lossless compressor} 
for strings of length $n$
from a finite alphabet $A$ is an injective function
$f_n:A^n\to\{0,1\}^*$, where $\{0,1\}^*=\{\emptyset,
0,1,00,01,10,\ldots\}$ is the set of all finite-length
binary strings. 
We call $n$ the {\em blocklength}
of $f_n$, and we also refer to $f_n$ as a {\em code}.

Let $x^n=(x_1,\ldots,x_n)\in A^n$ denote a
string of length $n$ from $A$.
The code $f_n$ is {\em prefix-free} if $f_n(x^n)$
is not a prefix of $f_n(y^n)$ whenever
$x^n\neq y^n$. The {\em description length}
of $x^n$ under a compressor $f_n$ is 
$\ell(f_n(x^n))$ bits, where $\ell(c)$ denotes the length
of a binary string $c\in\{0,1\}^*$.

A {\em source} $\Xp=\{X_n\;;\;n\geq1\}$ 
is an arbitrary random process with values in an alphabet~$A$.
The problem of understanding
and evaluating the best achievable performance 
of lossless compressors $f_n$ on strings $X^n=(X_1,\ldots,X_n)$
generated by some source $\Xp$
is often naturally examined in terms
of the fundamental underlying trade-off: 
We wish to have a small {\em compression
rate} $R>0$ while
also keeping the {\em excess-rate probability},
$\BBP(\ell(f_n(X^n)) >nR)$, small.

Formally, for each $\epsilon\in[0,1)$ and
each blocklength $n\geq 1$, the best
achievable rate with excess-rate
probability no greater than $\epsilon$ is 
\bqq
R^*(n,\epsilon)=\inf\big\{R>0:\inf_{f_n}
\BBP(\ell(f_n(X^n)) >nR)\leq\epsilon\big\},
\label{eq:Rstar}
\eqq
where the infimum is over all variable-length
compressors $f_n$.
Similarly, the best achievable excess-rate
probability at a given rate $R>0$
and blocklength $n$ is
\bqq
\epsilon^*(n,R)=\inf_{f_n}
\BBP(\ell(f_n(X^n)) >nR).
\label{eq:epsilonstar}
\eqq
As noted in~\cite{kontoyiannis-verdu:14},
the infima that appear in~(\ref{eq:Rstar})
and~(\ref{eq:epsilonstar}) are achieved by 
an optimal compressor $f_n^*$ which
is independent of the target rate $R$.

For prefix-free codes, the corresponding
fundamental limits $R^{\sf p}(n,\epsilon)$
and $\epsilon^{\sf p}(n,R)$ are defined in exactly
the same way, with the infima in~(\ref{eq:Rstar})
and~(\ref{eq:epsilonstar}) taken over the class
of all prefix-free compressors $f_n$. Although optimal
injective compressors are quite different from
optimal prefix-free codes, their performance
is tightly linked: For any source $\Xp$, any $R>0$,
and any $n\geq 1$, we have
\bqq
\epsilon^{\sf p}\Big(n,R+\frac{1}{n}\Big)
\leq\epsilon^*(n,R)
\leq\epsilon^{\sf p}(n,R);
\label{eq:epsilonstarp}
\eqq
see~\cite[Theorem~1]{kontoyiannis-verdu:14}. 

\subsection{Asymptotic approximations}

As it is virtually impossible to exactly evaluate 
the fundamental limits 
$R^*(n,\epsilon)$ and $\epsilon^*(n,R)$ in general,
most of the theoretical work in source coding has been 
concerned with
developing asymptotic approximations for various
classes of sources: Asymptotic expansions
are developed for $R^*(n,\epsilon)$ or $\epsilon^*(n,R)$
as the blocklength $n\to\infty$, and the leading terms
of these expansions are used as approximations.

The first-order behavior of the optimal rate $R^*(n,\epsilon)$ is determined
by the {\em entropy rate} $H(\Xp)$ of the source~$\Xp$:
The Shannon-McMillan theorem~\cite{shannon:48,mcmillan} 
implies that, for any stationary
and ergodic source $\Xp$ and any $\epsilon\in(0,1)$, 
we have, as $n\to\infty$:
\bqq
nR^*(n,\epsilon)=nH(\Xp)+o(n).
\label{eq:shannon}
\eqq
For memoryless sources,
the expansion~(\ref{eq:shannon}) can be 
refined~\cite{yushkevich,strassen:64b,kontoyiannis-verdu:14}
to
\be
nR^*(n,\epsilon)=nH(\Xp)+\sigma(\Xp)Q^{-1}(\epsilon)\sqrt{n}
-\frac{1}{2}\log n
+O(1),
\label{eq:strassen}
\ee
where
$\sigma^2(\Xp)$ is the 
source {\em varentropy}~\cite{kontoyiannis-verdu:13}
or {\em minimal coding variance}~\cite{kontoyiannis-97},
and $Q(z)=1-\Phi(z)$, $z\in\RL$, is the standard Gaussian 
tail function.
[Throughout, 
$\log$ denotes the logarithm to base~2,
and all information-theoretic functionals
are expressed in bits.]
When the excess-rate probability is required
to be very small, 
the optimal rate admits a different 
characterization~\cite{theocharous:toappear}:
For any memoryless
source $\Xp$ with marginal 
probability mass function (p.m.f.) 
$P$ on $A$ we have
$$
nR^*(n,2^{-n\delta})=nH(P_{\alpha^*})-\frac{1}{2(1-\alpha^*)}\log n +O(1),
$$
for any $\delta>0$ in an appropriate range, 
where, for $\alpha\in(0,1)$, the p.m.f.
\be
P_\alpha(x)=Z^{-1}P(x)^\alpha,
\quad x\in A,
\label{eq:Palpha}
\ee
with $Z=\sum_{y\in A}P(y)^\alpha$,
and $\alpha^*$ satisfies $D(P_{\alpha^*}\|P)=\delta$.
As usual, $H(P)$ denotes the entropy of a p.m.f.\ $P$
and $D(P\|Q)$ denotes the relative entropy between two
p.m.f.s $P$ and $Q$.

A corresponding series of results has been 
developed for the optimal excess-rate probability
$\epsilon^*(n,R)$,
when $\Xp$ is a memoryless source with 
marginal p.m.f.\
$P$ on $A$.
In the large-deviations regime,
for any rate $H(P)<R<\log|A|$,
$\epsilon^*(n,R)$ 
decays exponentially fast
with exponent given by $D(P_{\alpha^*}\|P)$,
where $\alpha^*$ satisfies
$H(P_{\alpha^*})=R$~\cite{dobrushin:62,jelinek:book,blahut:74}.
For fixed-length compressors,
this result was refined in~\cite{csiszar:71},
where
the exact polynomial pre-factor
of $\epsilon^*(n,R)$ was computed.
And in the moderate-deviations
regime, a different expansion 
for $\epsilon^*(n,R_n)$ 
was derived in~\cite{altug-wagner-K:13}
for rates $R_n$ close to the entropy,
$R_n=H(P)-c/\sqrt{n}$
for some constant~$c$.

\subsection{Sample complexity} 
\label{s:intronstar}

The results described above rely 
on {\em asymptotic} arguments,
based on careful
examination of the behavior,
as the blocklength $n\to\infty$, 
of the {\em information} functional,
$-\log P_n(X^n),$
where $P_n$ denotes the p.m.f.\ of $X^n$ on $A^n$. 
Increasingly accurate expressions are developed
by taking $n\to\infty$ and 
applying
the law of large numbers, the central limit theorem,
and asymptotic estimates obtained from 
large- and moderate-deviations bounds.

We introduce a different,
{\em genuinely non-asymptotic} approach
to quantifying the fundamental performance limits
of lossless data compression. This approach
is partly motivated by parallel developments
in the statistics and the computer science 
literature, as outlined in
Section~\ref{s:history}.

Since the goal of data compression is to 
select codes such that both the rate $R$ 
and the excess-rate probability can be made 
small enough to satisfy given design requirements,
we define the {\em sample complexity} $n^*$ 
as the shortest blocklength at which such codes exist.
Specifically, for an arbitrary source
$\Xp=\{X_n\;;\;n\geq 1\}$ 
on a finite alphabet $A$, and 
for any $\epsilon\in(0,1)$
the {\em sample complexity}
$n^*(\Xp,\epsilon)$ is defined as,
$$n^*(\Xp,\epsilon)
=
	\inf\left\{n\geq 1\;:\;
	\inf_{f_n}\BBP(\ell(f_n(X^n))>nR)\leq \epsilon
	\;\mbox{and}\;\frac{2^{nR}}{|A|^n}\leq\epsilon,
	\;\mbox{for some}\;R>0\right\},
$$
or, more compactly,
\bqq
n^*(\Xp,\epsilon)
=
	\inf\left\{n\geq 1\;:\;
	\inf_{f_n,R>0}\max\left\{
	\BBP(\ell(f_n(X^n))>nR),
	\frac{2^{nR}}{|A|^n}\right\}\leq\epsilon
	\right\},
	\label{eq:Nstargeneral}
\eqq
where the infimum in~(\ref{eq:Nstargeneral})
is over all variable-length codes $f_n$
and all positive rates $R$.

\newpage

The point of the definition~(\ref{eq:Nstargeneral}) is simple:
Knowing the value of $n^*(\Xp,\epsilon)$ -- 
or having good bounds on it -- clearly tells
the practitioner exactly how large the blocklength
$n$ needs to be taken so that
explicit performance guarantees can be provided
for both 
the excess-rate probability
and the compression performance.

Note that the form in which the rate appears --
namely, as the ratio $2^{nR}/|A|^n$ -- is chosen so that
the rate $R$ and the excess rate probability can be considered
at the same scale. Further motivation and interpretation
for the exact form of the definition of $n^*(\Xp,\epsilon)$
and its operational consequences for compression
are given in Sections~\ref{s:FLIID} and~\ref{s:VLIID}.

\subsection{Outline of main results}

Section~\ref{s:preliminary} contains the 
definitions, notation and terminology used throughout 
the paper, along with the statements of two basic known 
results needed later.

In Section~\ref{s:FLIID} we consider
the simpler problem of determining 
the sample complexity of {\em fixed-length}
compressors. This allows for the presentation and interpretation
of the main ideas in this work clearly, in the least
technical setting.

The sample complexity of variable-length compression
is considered in detail in 
Section~\ref{s:VLIID}.
First, a number of properties
of the fundamental limit $n^*(\Xp,\epsilon)$
defined in~(\ref{eq:Nstargeneral})
are derived, including
general relationships which show
that the sample complexity 
of variable-length compressors is tightly coupled
with the sample complexity of prefix-free codes
and fixed-length codes (Theorems~\ref{thm:fltostar} 
and~\ref{thm:vltopf}).
Then the sample complexity is evaluated
in the case of memoryless sources~$\Xp$.
Theorem~\ref{thm:NstarIIDvl}
states that
\bqq
n^*(\Xp,\epsilon)\asymp\frac{\log(1/\epsilon)}
{D_{1/2}(P\|U)},
\label{eq:preview}
\eqq
where $P$ is the marginal source distribution, 
$U$ is the uniform p.m.f.\ on the same alphabet as~$\Xp$,
$D_{1/2}(P\|Q)$ denotes the R\'{e}nyi divergence
of order 1/2 between two p.m.f.s $P,Q$ on the same alphabet,
and where for
two nonnegative expressions $f$ and $g$ we write
$f\asymp g$ to signify that there are
absolute positive constants $C,C'$ such that $Cg\leq f\leq C'g$.
The operational interpretation of~(\ref{eq:preview})
in the context of data compression, as well as its relation to
the earlier Gaussian approximation~(\ref{eq:strassen}) is discussed
in Section~\ref{s:example}.

The expression for $n^*(\Xp,\epsilon)$ in~(\ref{eq:preview})
is non-asymptotic, it holds uniformly in $\epsilon$
and the distribution $P$, and it is very
simple. Moreover, the implied constants
of the upper and lower bounds in~(\ref{eq:preview})
are explicit and of very reasonable magnitude,
see~(\ref{eq:nstarexplicit1}) and~(\ref{eq:nstarexplicit2}).
For example, in~\eqref{eq:nstarexplicit2} we show that,
for $0<\epsilon<\min\{\frac{1}{8},\frac{1}{|A|}\},$
$$\frac{\log(1/\epsilon)}{4D_{1/2}(P\|U)}
\leq n^*(\Xp,\epsilon)\leq
\frac{3\log(1/\epsilon)}{D_{1/2}(P\|U)}.$$
Interestingly, the key property of the source
that determines $n^*(\Xp,\epsilon)$ is not its entropy
$H(P)$ but its R\'{e}nyi entropy $H_{1/2}(P)$ of order 1/2, 
since we always have
$D_{1/2}(P\|U)=\log |A|-H_{1/2}(P)$.

The main idea in the proof of~(\ref{eq:preview}) is that
$n^*(\Xp,\epsilon)$ 
can be easily related 
to the quantity
$$N^{\sf fl}(\Xp,\epsilon)
=
	\inf\left\{n\geq 1\;:\;
	\inf_{C_n\subset A^n}\left[
	P^n(C_n^c)+
	\frac{|C_n|}{|A|^n}\right]\leq\epsilon
	\right\},
$$
defined below in~(\ref{eq:Nfl}).
It is not hard to see that $N^{\sf fl}(\Xp,\epsilon)$
is the sample complexity of the simple-versus-simple
hypothesis test between the marginal
source p.m.f.\ $P$ and the uniform p.m.f.\ $U$.
Then~(\ref{eq:preview})
follows by translating known~\cite{bar-yossef:02,canonne:22}
sample complexity bounds for hypothesis testing
to the present setting.

In Section~\ref{s:markov} we examine the sample complexity
of Markov sources. Theorem~\ref{thm:irreducible} gives upper 
and lower bounds to $n^*(\Xp,\epsilon)$ for an arbitrary 
irreducible Markov source $\Xp$ on a finite alphabet,
but these are more involved than~(\ref{eq:preview}).
In particular, they depend on the initial distribution
of the chain and on the spectral properties of a matrix
associated with its transition matrix. Cleaner bounds
are obtained in Theorem~\ref{samplecomplexitysymmetric}
for the special class of symmetric Markov sources.
In both cases, the main property of the source that
determines its sample complexity is the
R\'{e}nyi divergence {\em rate} $D_{1/2}(\Xp\|\Up)$
between $\Xp$ and the independent and identically
distributed (i.i.d.) uniform source $\Up$.

Finally, in Section~\ref{s:universal} we consider
the problem of {\em universal} data compression
for general classes
of memoryless sources. The gist of our
approach is to relate this problem
to {\em composite}
hypothesis testing,
specifically to
{\em identity testing}. This connection is
discussed further
in Section~\ref{s:history}
below. 

Let $\clQ$ be
an arbitrary collection of p.m.f.s on an
alphabet $A$ of size $m=|A|$. We define the {\em universal sample complexity}
$n^*(\clQ,\epsilon)$ of $\clQ$
as the shortest blocklength $n$ for which
there is a variable-length compressor that achieves
an excess-rate probability no more than $\epsilon$
on {\em every} memoryless source with marginal p.m.f.\
$P\in\clQ$, at some rate $R$ such that $2^{nR}/|A|^n\leq\epsilon$.
Upper and lower bounds on $n^*(\clQ,\epsilon)$
are derived for several different classes of families $\clQ$,
including the special class of $\clQ_\delta$ defined as:
$$\clQ_\delta=\big\{P:D_{1/2}(P\|U)\geq \delta\big\}.$$
For any family $\clQ$ let
 $D_{1/2}(\clQ\|U)=\inf_{P\in \clQ}D_{1/2}(P\|U)$,
so that $D_{1/2}(\clQ_\delta\|U)=\delta$.
In Theorem~\ref{thm:canonne} we show that
\bqq
n^*(\clQ_\delta,\epsilon)\leq C\frac{\sqrt{m}
\big[\log(1/\epsilon)+\frac{1}{18}\big]}{\delta},
\label{eq:separation}
\eqq
along with a corresponding lower bound
also expressed in terms of $\log(1/\epsilon)$,
$\delta$ and $m$.
These results imply that the universal sample complexity
of the family $\clQ_\delta$ is determined
by its $D_{1/2}$-distance,
$\delta=D_{1/2}(\clQ_\delta\|U)$, from the uniform.

Following the same path as the parallel development 
of statistical ideas, we can solve these 
sample complexity bounds 
for $\delta$ to obtain corresponding
bounds for the optimal {\em separation distance} $\delta^*(n,\epsilon)$
at a given blocklength $n$.
For example, in the moderate confidence regime 
where~$\epsilon$ is restricted in a region bounded away from
zero (e.g., taking $\epsilon$ fixed),
Theorem~\ref{thm:canonne} implies that
$$\delta^*(n,\epsilon)\asymp
\frac{\sqrt{m}}{n},
$$
which leads to a ``separation rates'' interpretation
of the bound in~(\ref{eq:separation}):
At blocklength $n$, the largest family of memoryless sources
that can be compressed with excess-rate probability 
bounded by $\epsilon$ at a rate $R$ bounded 
as $2^{nR}/|A|^n\leq\epsilon$, is necessarily separated
from the uniform
by a $D_{1/2}$-distance $\asymp\sqrt{m}/n$.

\subsection{History and general remarks on sample complexity}
\label{s:history}

The deep connections between hypothesis testing and lossless
data compression were identified and explored early on,
see, e.g.,~\cite{csiszar:book,csiszar-shields:04}.

In statistics, the classical paradigm for hypothesis
testing was set by Neyman and Pearson 
in 1933~\cite{neyman:33}: Fix the type-I error
at some level $\epsilon$, and minimize the type-II error
among all tests of that size.
The first conceptual departure from this 
came in Wald's decision-theoretic work~\cite{wald:45,wald:book},
where minimizing the sum of the errors over all tests
is viewed as a risk minimization problem.
The idea of sample complexity -- namely, the
smallest sample size at which the sum of the errors
or, essentially equivalently, their maximum, can be made smaller
than $\epsilon$ -- was first advocated by
Le Cam~\cite{lecam:56, lecam:60b}.
The modern minimax formulation of
nonparametric hypothesis testing, 
along with the associated study
of optimal separation rates, is primarily due
to Ingster~\cite{ingster:84,ingster:85},
see also~\cite{ingster:book}.
Over the past 20 years, these ideas
have also been adopted in problems of distribution 
testing over discrete spaces, with an emphasis
on non-asymptotic bounds. That literature
includes a number of results
relevant to the present work;
see the reviews~\cite{canonne:20,canonne:22b}
and the references below.

Historically, almost all core information-theoretic problems
have been stated and treated in a setting analogous
to the Neyman-Pearson framework. For example,
in channel~coding, the error probability is
fixed and the communication rate is maximized
over all codes that satisfy the error probability
constraint. Fundamental limits are subsequently
characterized via asymptotically tight approximations
as the blocklength $n\to\infty$~\cite{cover:book2,gao:26}.

Since the late 1990s, a number of authors --
including Jacob Ziv in his 1997 Shannon Lecture~\cite{ziv:97} --
have advocated that the focus be shifted
to non-asymptotic results.
In this work, we propose that the 
sample complexity formulation provides
a useful framework within which
the core classical information-theoretic
problems can be recast, and where
powerful and informative non-asymptotic bounds 
to fundamental performance
limits can naturally 
be established.

The connection between lossless data compression
and the sample complexity 
of hypothesis testing problems
goes well beyond merely the problem formulation.
The following works contain results related
to the bounds
developed in this paper. 
For memoryless sources, the results
in Theorem~\ref{thm:NstarIIDvl}
follow from the
bounds in~\cite{bar-yossef:02,canonne:22}.
For Markov sources, the work closest in spirit
to our results in Section~\ref{s:markov} is
reported in~\cite{wang:23}.
Earlier work in~\cite{wolfer:20,wolfer:21} 
and~\cite{chan:21} involves sample complexity
bounds in terms of $L_1$ distance,
and testing between symmetric Markov chains 
is considered 
in~\cite{daskalakis:18} and~\cite{cherapanamjeri:19}.
The problem of universal data compression
as formulated in Section~\ref{s:universal}
is closely related to 
{\em identity testing} and {\em goodness-of-fit tests}.
Early relevant work 
includes~\cite{paninski:08,goldreich:11,valiant:14,acharya:15,diakonikolas:15},
with bounds of various forms also proved 
in~\cite{diakonikolas:19,diakonikolas:19b,acharya:20,acharya:21}.
Some of the strongest such bounds that are also
closely related to our development are
established in~\cite{diakonikolas:18}.

\section{Preliminaries}
\label{s:preliminary}

We begin with some general definitions and assumptions that
remain in effect throughout the paper.

A {\em source} $\Xp=\{X_n\;;\;n\geq 1\}$ is an arbitrary sequence
of random variables $X_n$ with values in a common finite
alphabet $A=\{a_1,\ldots,a_m\}$ of size $|A|=m$.
For each $n\geq 1$, the probability mass function (p.m.f.)
of $X^n=(X_1,\ldots,X_n)$
on $A^n$ is denoted by $P_n$, so that
$P_n(x^n)=\BBP(X^n=x^n)$, $x^n\in A^n$.
We identify the p.m.f.\ $P_n$ with the 
probability measure it induces on $A^n$ and
we write, for example, 
$P_n(C_n)$ for the probability $\BBP(X^n\in C_n)$,
when $C_n$ is a subset of $A^n$.

For any two nonnegative expressions $f,g$ that 
may depend on one or several variables,
$x,y$ and $z$ for example, 
we write, 
$$f(x,y,z)\asymp g(x,y,z),$$ to signify that
$C g(x,y,z)\leq f(x,y,z)\leq C' g(x,y,z)$
for some finite, positive constants $C,C'$
and for all values of $x,y,z$ considered.

The entropy $H(P)$ of a p.m.f.\ $P$ on a finite alphabet
$B$ is defined as usual by 
$$H(P)=-\sum_{x\in B}P(x)\log P(x),$$
where $\log$ denotes the base-2 logarithm. The relative
entropy between two p.m.f.s $P,Q$ on the same alphabet $B$
is 
$$D(P\|Q)=\sum_{x\in B}P(x)\log\Big(\frac{P(x)}{Q(x)}\Big).$$
The R\'{e}nyi entropy of order $\alpha\in(0,1)$ of a p.m.f.\
$P$ on $B$ is
$$H_{\alpha}(P)=\frac{1}{1-\alpha}\log\left(\sum_{x\in B}P(x)^\alpha\right),$$
and the 
R\'{e}nyi divergence of order $\alpha$ between two p.m.f.s
$P,Q$ on $B$ is
$$D_{\alpha}(P\|Q)
=
	\frac{1}{\alpha-1}\log
	\left(\sum_{x\in B}P(x)^{\alpha}Q(x)^{1-\alpha}
	\right).
$$
When $\alpha=1/2$, for
any two p.m.f.s $P,Q$, we always have~\cite{vanerven:14},
\bqq
D_{1/2}(P\|Q)\leq D(P\|Q).
\label{eq:DhalfD}
\eqq
And if $U$ is the uniform p.m.f.\ on $B$ then 
$$D_{1/2}(P\|U)=\log|B|-H_{1/2}(P).$$

Clearly $D_{1/2}(P\|Q)$ is closely related to the Hellinger distance
$\clH_2(P,Q)$, given by
\begin{align*}
\clH^2_2(P,Q)
&=
	\frac{1}{2}\sum_{x\in B}\big(\sqrt{P(x)}-\sqrt{Q(x)}\big)^2\\
&=
	1-\sum_{x\in B}\sqrt{P(x)Q(x)},
\end{align*}
so that
$$D_{1/2}(P\|Q)=-2\log \big(1-\clH_2^2(P,Q)\big).$$
The {\em R\'{e}nyi divergence rate} of order $1/2$ between
two sources $\Xp$ and $\Yp$ on the same alphabet $A$,
and with marginals $\{P_n\}$ and $\{Q_n\}$, respectively,
is defined by
$$D_{1/2}(\Xp\|\Yp)=\lim_{n\to\infty}\frac{1}{n}D_{1/2}(P_n\|Q_n),$$
whenever the limit exists.
Finally, the total variation distance between $P$ and $Q$ is,
as usual, defined by:
$$\|P-Q\|_{\rm TV}
=\sup_{C\subset B}|P(C)-Q(C)|
=\sup_{C\subset B}\big(P(C)-Q(C)\big)
=\frac{1}{2}\sum_{x\in B}
|P(x)-Q(x)|.
$$

We will need two standard properties of R\'{e}nyi 
divergence. The first one is a special case of
its tensorization property;
see, e.g.,~\cite{vanerven:14}.

\begin{proposition}[Tensorization of $D_\alpha(P\|Q)$]
\label{prop:tensorize}
Suppose $P,Q$ are arbitrary p.m.f.s on a finite alphabet $A$.
Then, for any $n\geq 1$ and any $\alpha\in(0,1)$,
$$D_\alpha(P^n\|Q^n)=nD_\alpha(P\|Q),$$
where $P^n,Q^n$ denote the corresponding product
p.m.f.s on $A^n$.
\end{proposition}

The following proposition states a pair of well-known
inequalities that relate $D_{1/2}(P\|Q)$ to total variation,
see, e.g.,~\cite{lecam:86}.
Since they 
are usually stated in terms of Hellinger 
distance rather than R\'{e}nyi divergence,
we prove Proposition~\ref{prop:DTV}
in Appendix~\ref{app:DTV}
for the sake of completeness.

\begin{proposition}[R\'{e}nyi divergence and total variation]
\label{prop:DTV}
For any pair of p.m.f.s $P,Q$ on the same alphabet $B$:
$$2^{-D_{1/2}(P\|Q)-1} \leq 1 - \|P-Q\|_{\rm TV} 
\leq 2^{-\frac{1}{2}D_{1/2}(P\|Q)}.$$
\end{proposition}

\section{Sample complexity of fixed-length compression}
\label{s:FLIID}

In order to present the key ideas as clearly as possible,
we first examine the simpler class of {\em fixed-length} compressors.
In this case, and assuming the source distribution is known,
the form of the sample complexity is more 
straightforward 
to motivate and interpret, the essential bounds
are easy to establish, and the connection with 
hypothesis testing is explicit.

\subsection{Fixed-length codes and sample complexity}
\label{s:FLgeneral}

A {\em fixed-length lossless compressor} for strings of
length $n$ from a finite alphabet $A$, is fully specified
by a {\em codebook} $C_n\subset A^n$:
If $x^n\in C_n$, then the encoder describes $x^n$ by 
describing its index in $C_n$, using
$\lceil\log|C_n|\rceil$ bits; otherwise,
it declares an error.
When the string $X^n$ to be compressed is generated by 
some source $\Xp=\{X_n\;;\;n\geq 1\}$,
the goal is to achieve good compression by selecting
a codebook with small size $|C_n|$,
while also keeping its
{\em error probability}, 
$\BBP(X^n\not\in C_n)=P_n(C_n^c)$,
small.

Therefore, we define the sample complexity 
of fixed-length compression
as the shortest blocklength $n$ at which there is a codebook
$C_n$ with appropriately small size and small error probability.
Specifically, for any $\epsilon\in(0,1)$, 
the {\em fixed-length sample complexity}
$n^{\sf fl}(\Xp,\epsilon)$ of the source $\Xp$
is the smallest $n$ such that
the error probability $P_n(C_n^c)$ {\em and}
the proportion of strings $x^n$ that belong 
to the codebook $C_n$ can {\em both} be made
smaller than $\epsilon$:
$$n^{\sf fl}(\Xp,\epsilon)
=\inf\left\{n\geq 1\;:\;\inf_{C_n\subset A^n}\max
\left\{P_n(C_n^c),\frac{|C_n|}{|A|^n}\right\}
\leq \epsilon\right\}.
$$
We will also find it convenient to work with the related
quantity $N^{\sf fl}(\Xp,\epsilon)$, defined similarly
but with the maximum replaced by a sum. For any $\epsilon\in(0,2)$:
\bqq
N^{\sf fl}(\Xp,\epsilon)
=\inf\left\{n\geq 1\;:\;\inf_{C_n\subset A^n}
\left[P_n(C_n^c)+\frac{|C_n|}{|A|^n}\right]
\leq \epsilon\right\}.
\label{eq:Nfl}
\eqq

It is immediate from the definitions that
for any source $\Xp$ and any $\epsilon\in(0,1)$:
\bqq
N^{\sf fl}(\Xp,2\epsilon)
\leq n^{\sf fl}(\Xp,\epsilon)
\leq N^{\sf fl}(\Xp,\epsilon).
\label{eq:Nflnfl}
\eqq
Therefore, results about $n^{\sf fl}$ readily translate
to results about $N^{\sf fl}$ and vice versa.

The reason why it is often easier to work with $N^{\sf fl}$
rather than with $n^{\sf fl}$ is because $N^{\sf fl}$ admits
a simpler representation in terms of total variation.
The following observation is a version of a result
known as Le Cam's lemma, c.f.~\cite{lecam:86}.

\begin{proposition}[Le Cam's lemma]
\label{prop:lecam}
Let $U$ denote the uniform p.m.f.\ on $A$.
For any source $\Xp$ on $A$ and 
any $\epsilon\in(0,2)$:
$$N^{\sf fl}(\Xp,\epsilon)
=\inf\big\{n\geq 1\;:\;1-\|P_n-U^n\|_{\rm TV}
\leq \epsilon\big\}.
$$
\end{proposition}

\noindent
{\sc Proof.}
Since $U$ is uniform, we have
\begin{align*}
\inf_{C_n}\left[P_n(C_n^c)+\frac{|C_n|}{|A|^n}\right]
&=
	\inf_{C_n}\big[P_n(C_n^c)+U^n(C_n)\big]\\
&=
	1-\sup_{C_n}\big[P_n(C_n)-U^n(C_n)\big]\\
&=
	1-\|P_n-U^n\|_{\rm TV},
\end{align*}
and the result follows from the definition of $N^{\sf fl}(\Xp,\epsilon)$.
\qed

Proposition~\ref{prop:lecam} is the starting point
for many of the bounds derived in this paper.

\subsection{Memoryless sources}
\label{s:IID}

Our first sample complexity result says that, if $\Xp$
is a memoryless source with distribution $P$, then:
$$n^{\sf fl}(\Xp,\epsilon)
\asymp\frac{\log(1/\epsilon)}{D_{1/2}(P\|U)}.$$

\begin{theorem}[Fixed-length sample complexity of memoryless
sources]
\label{thm:nfl}
$\;$ Let $U$ denote the uniform p.m.f.\ on $A$.
For any memoryless source $\Xp$ with marginal p.m.f.\ 
$P\neq U$
on $A$
and any $\epsilon\in(0,1)$, 
the fixed-length sample complexity of $\Xp$ satisfies:
\bqq
\frac{\log ( 1/\epsilon )-2}{D_{1/2}(P\|U)} 
\leq n^{\sf fl}(\Xp,\epsilon) 
\leq \frac{ 2\log( 1/\epsilon )}{D_{1/2}(P\|U)}+1.
\label{eq:nfl1}
\eqq
In particular, for $0<\epsilon<\min\{\frac{1}{8},\frac{1}{m}\}$,
\bqq
\frac{\log ( 1/\epsilon )}{3D_{1/2}(P\|U)} 
\leq n^{\sf fl}(\Xp,\epsilon) 
\leq \frac{ 3\log( 1/\epsilon )}{D_{1/2}(P\|U)}.
\label{eq:nfl2}
\eqq
\end{theorem}

Before giving the proof, some important
remarks are in order.

\medskip

\noindent
{\bf Remarks.}
\begin{enumerate}
\item {\em Hypothesis testing.}
The computation in the proof of Proposition~\ref{prop:lecam}
shows that
$N^{\sf fl}(\Xp,\epsilon)$ can be expressed as,
$$N^{\sf fl}(\Xp,\epsilon)
=\inf\Big\{n\geq 1:\inf_{C_n\subset A^n}\big[P^n(C^c_n)+U^n(C_n)\big]
\leq \epsilon\Big\}.$$
This is exactly the sample complexity of the hypothesis
test ``$P$ versus $U$,''
with $P^n(C_n^c)$ and $U^n(C_n)$ being the two error probabilities
associated with a decision region $C_n$.
This highlights the connection between
lossless data compression and hypothesis testing,
at the level of sample complexity.
\item {\em Proof.}
Once the data compression question is formulated
in terms of $N^{\sf fl}$ and the connection between
$n^{\sf fl}$ and $N^{\sf fl}$ is identified, Theorem~\ref{thm:nfl}
immediately follows from the corresponding hypothesis
testing bounds, see, e.g., \cite{bar-yossef:02,canonne:22}.
The proof given below is a slightly streamlined version
of these earlier results, using $D_{1/2}(P\|U)$ in place
of the Hellinger distance.
\item {\em R\'{e}nyi entropy determines sample complexity.}
The upper and lower bounds in~(\ref{eq:nfl1})
and~(\ref{eq:nfl2}) can be viewed as achievability
and converse results, respectively.
A more detailed discussion of this point is given in 
Section~\ref{s:example}.
Importantly, the key information-theoretic
functional that determines the behavior
of the fundamental limit $n^{\sf fl}(\Xp,\epsilon)$ 
is not
the entropy $H(P)$ of the source distribution $P$, 
but its R\'{e}nyi divergence 
of order $1/2$ to the uniform distribution,
$D_{1/2}(P\|U)$. Or, equivalently,
the R\'{e}nyi entropy $H_{1/2}(P)$, since
$D_{1/2}(P\|U)=\log|A|-H_{1/2}(P)$.
\item {\em Exponential behavior in $n$.}
Solving~(\ref{eq:nfl2}) for $\epsilon$ says
that, at best, both $|C_n|/|A|^n$ and the
error probability behave like,
\bqq
2^{-n\Theta(D_{1/2}(P\|U))}.
\label{eq:army1}
\eqq
Therefore, the optimal error probability
decays (as expected) exponentially fast with
$n$, and the rate that optimally balances
the error probability is essentially
determined 
by $D_{1/2}(P\|U)$. It is perhaps
worth noting that such little
effort readily establishes the exponential
decay of error probability in this setting.
\item {\em Chernoff information.}
From the proof of Theorem~\ref{thm:nfl}
it follows that,
regardless of the value of $\epsilon$,
the optimal fixed-length compressor
corresponds to the codebook $C_n$ that
achieves the supremum in the definition
of the total variation distance between
$P^n$ and $U^n$, which is
$$
C_n=\{x^n\in A^n:P^n(x^n)\geq U^n(x^n)\}.$$
For this codebook it is easy to
compute the exponential behavior
of both $|C_n|/|A|^n$ and $P^n(C_n^c)$.
Indeed, to first order in the exponent,
both of these behave like
\bqq
2^{-n\clC(P,U)},
\label{eq:army2}
\eqq
where $\clC(P,U)$ is the {\em Chernoff information}
between $P$ and $U$, given by
\be
\clC(P,U)=\inf \big\{D(Q\|P)\;:\;\mbox{p.m.f.s $Q$ s.t.}\;
D(Q\|U)\leq D(Q\|P)\big\}.
\label{eq:chernoff}
\ee
This computation also offers another partial
explanation for why it is $D_{1/2}(P\|U)$ instead of the entropy
$H(P)$ that determines the sample complexity. 
Since the error probability here necessarily decreases
exponentially, we are no longer in ``Shannon regime''
where the rate is optimized while keeping the excess-rate
probability bounded above by a fixed $\epsilon>0$.
Instead, we are in the error-exponents regime
where, in the present balanced case, the optimal
exponent is given by $\clC(P,U)$.
\item {\em Solidarity.}
Examining the optimal behaviour of
$|C_n|/|A|^n$ and of 
the error probability $P^n(C_n^c)$, 
in~(\ref{eq:army1}) we showed that they
both behave like $\approx 2^{-n\Theta(D_{1/2}(P\|U))}$,
while in~(\ref{eq:army2}) we claim that there
are $\approx 2^{-n\clC(P,U)}$.
The reconciliation of these seemingly 
different results
comes from the fact that $\clC(P,U)\asymp D_{1/2}(P\|U)$.
In fact, it is not hard to show that for any source
distribution $P$:
$$\frac{1}{2}D_{1/2}(P\|U)\leq \clC(P,U)\leq D_{1/2}(P\|U).$$
This also shows that the bounds in Theorem~\ref{thm:nfl}
could have equivalently been stated in terms of 
$\clC(P,U)$ instead of $D_{1/2}(P\|U)$. 
We choose $D_{1/2}(P\|U)$ instead of $\clC(P,U)$ because
of its more explicit form and the more natural way in which
it arises in the proof.

\item
{\em More general criteria.}
One may reasonably wish to define
a richer version of sample complexity,
where the rate 
is constrained differently
from the error probability.
For example, 
in the present setting of fixed-length
compression, it is reasonable to consider,
for all $\epsilon_1,\epsilon_2\in(0,1)$, 
the following more
general version of sample complexity,
$$n^{\sf fl}(\Xp,\epsilon_1,\epsilon_2)
=\inf\left\{n\geq 1\;:\;
P_n(C_n^c)\leq \epsilon_1\;\mbox{and}\;
\frac{|C_n|}{|A|^n}\leq \epsilon_2,
\;\mbox{for some}\;C_n\subset A^n\right\},
$$
or the essentially equivalent ``weighted'' version,
for $\epsilon>0$ and $\pi\in(0,1)$,
$$
N^{\sf fl}_\pi(\Xp,\epsilon)
=\inf\left\{n\geq 1\;:\;
\inf_{C_n\subset A^n}\left[
\pi P_n(C_n^c)+(1-\pi)
\frac{|C_n|}{|A|^n}\right]\leq \epsilon \right\}.
$$
In the context of hypothesis testing,
this extension has recently been carried
out in~\cite{pensia:24}, 
where it is shown that the
behaviour of 
$N_\pi^{\sf fl}(\Xp,\epsilon)$
can be classified into different regimes,
depending on the relative values of $\pi$
and $\epsilon$. For example, when $\epsilon$
is constrained to be bounded away from zero
and $\epsilon<\pi/4$, then a simple adaptation
of the proofs in~\cite{pensia:24} shows that
$N_\pi^{\sf fl}(\Xp,\epsilon)\asymp\frac{\log(1/\epsilon)}{D_\lambda(P\|U)}$,
for an appropriate value of $\lambda\neq 1/2$
that depends on $\pi$.

On the other
hand, in the more interesting regime when 
$\epsilon$ can be taken arbitrarily close to zero
(so that the bounds blow up), we still have, for any $\pi\in(0,1/2]$,
$$N_\pi^{\sf fl}(\Xp,\epsilon)\asymp\frac{\log(1/\epsilon)}{D_{1/2}(P\|U)},$$
uniformly over all $\epsilon\in(0,\pi^2]$.
Once again, this offers evidence that the R\'{e}nyi entropy
$D_{1/2}(P\|U)$ is the ``right'' functional of the
source distribution in the context of evaluating the sample
complexity of data compression.

Although the results in~\cite{pensia:24} can be
further explored to obtain 
corresponding bounds for data compression,
we will not pursue this further in this paper.

\item {\em Uniform sources.} Strictly speaking,
the results of the theorem remain valid as stated in the
case when $P$ is the uniform distribution over a finite
alphabet $A$, but in this case it is actually easy to evaluate 
$n^{\sf fl}(\Xp,\epsilon)$ and
$N^{\sf fl}(\Xp,\epsilon)$ precisely, directly 
from the definition. We have that
$N^{\sf fl}(\Xp,\epsilon)$ equals 1 if $\epsilon\geq 1$
and $\infty$ otherwise.
If $|A|$ is even, we similarly have that
$n^{\sf fl}(\Xp,\epsilon)$ equals 1 if $\epsilon\geq 1/2$
and $\infty$ otherwise. And if $|A|$ is odd, then
$$
n^{\sf fl}(\Xp,\epsilon)=\left\lceil\frac{-\log(2\epsilon-1)}{\log|A|}
\right\rceil,
\quad\mbox{for}\;1/2<\epsilon<1,
$$
and 
$n^{\sf fl}(\Xp,\epsilon)=\infty$ for $0<\epsilon\leq 1/2$.
\end{enumerate}

\noindent
{\sc Proof.}
We first establish analogous bounds for $N^{\sf fl}(\Xp,\epsilon)$.
Given $\epsilon\in(0,1)$ and $P$, let
$\epsilon(n)=1-\|P^n-U^n\|_{\rm TV}$.
Using the upper and lower bounds in Proposition~\ref{prop:DTV}
followed by the tensorization identity
in Proposition~\ref{prop:tensorize}, yields
$$2^{-nD_{1/2}(P\|U)-1} \leq \epsilon(n)
\leq 2^{-\frac{n}{2}D_{1/2}(P\|U)}.$$
And since, by Le Cam's lemma,
$\epsilon(N^{\sf fl}(\Xp,\epsilon)) \leq 
\epsilon 
< \epsilon(N^{\sf fl}(\Xp,\epsilon)-1)$,
we have,
$$\frac{\log ( 1/\epsilon )-1}{D_{1/2}(P\|U)} 
\leq N^{\sf fl}(\Xp,\epsilon) \leq 
\frac{ 2\log( 1/\epsilon )}{D_{1/2}(P\|U)}+1.$$
The bounds in~(\ref{eq:nfl1}) follow from the observation~(\ref{eq:Nflnfl}),
and~(\ref{eq:nfl2}) follows by direct calculation
and the fact that $D_{1/2}(P\|U)\leq\log |A|$.
\qed

In this section we obtained bounds on the sample complexity
of fixed-length data compression, by relating the 
evaluation of the fundamental limit $n^*(\Xp,\epsilon)$
directly to a problem in hypothesis testing.
We are now ready to move on to the more interesting,
somewhat less straightforward,
and more practically relevant problem of 
variable-length lossless compression.

\section{Sample complexity of variable-length compression}
\label{s:VLIID}

\subsection{Variable-length codes and sample complexity}
\label{s:VLgeneral}

Recalling the discussion in Section~\ref{s:intronstar},
for an arbitrary source $\Xp=\{X_n\;;\;n\geq 1\}$ on $A$
and any $\epsilon\in(0,1)$,
we define the {\em variable-length sample complexity}
$n^*(\Xp,\epsilon)$ of $\Xp$ as
\bqq
n^*(\Xp,\epsilon)
=\inf\left\{n\geq 1\;:\;
\inf_{f_n,R>0}\max\left\{\BBP\big(\ell(f_n(X^n))>nR\big),\;\frac{2^{nR}}{|A|^n}
\right\}\leq\epsilon\right\}.
\label{eq:nstar}
\eqq
As in the case of fixed-length compression, 
for $\epsilon\in(0,2)$ we also define:
\bqq
N^*(\Xp,\epsilon)=\inf\left\{n\geq 1\;:\;
\inf_{f_n,R>0}\left[\BBP\big(\ell(f_n(X^n))>nR\big)+\frac{2^{nR}}{|A|^n}
\right]\leq\epsilon\right\}.
\label{eq:Nstar}
\eqq
In both cases, the infimum is over all lossless compressors
$f_n$ on $A^n$ and all rates $R>0$.
Also, as for $n^{\sf fl}$ and $N^{\sf fl}$ earlier, 
again we always have,
\bqq
N^*(\Xp,2\epsilon)
\leq n^*(\Xp,\epsilon)
\leq N^*(\Xp,\epsilon).
\label{eq:Nstarnstar}
\eqq

The best achievable performance
of variable-length compressors
is very closely related to that of fixed-length
codes. 
The proof of Theorem~\ref{thm:fltostar}
is based on an explicit combinatorial construction
and elementary probabilistic bounds.

\begin{theorem}[Fixed- vs.\ variable-length sample complexity]
\label{thm:fltostar}
For any source $\Xp$ on $A$ and any $\epsilon\in(0,1)$,
\bqq
n^{\sf fl}(\Xp,2\epsilon) \leq n^*(\Xp,\epsilon)
\leq n^{\sf fl}(\Xp,\epsilon),
\label{eq:nfltonstar}
\eqq
and similarly,
\bqq
N^{\sf fl}(\Xp,2\epsilon) \leq N^*(\Xp,\epsilon)
\leq N^{\sf fl}(\Xp,\epsilon),
\label{eq:Nfltonstar}
\eqq
with the obvious understanding that 
$n^{\sf fl}(\Xp,\epsilon)=N^{\sf fl}(\Xp,\epsilon)=1$
for $\epsilon\geq 1$.
\end{theorem}

\noindent
{\sc Proof.}
We only prove~(\ref{eq:Nfltonstar}); the proof of~(\ref{eq:nfltonstar})
is similar.

For the upper bound, we note that, given any 
nonempty
$C_n\subset A^n$, there is a variable-length
compressor $f_n$ that maps the elements of $C_n$
to binary strings of length no larger than 
$k(C_n)=\lfloor\log|C_n|\rfloor$,
and all elements of $C_n^c$ to arbitrary
binary strings of length at least $k(C_n)+1$.
And for the case when $C_n$ is empty, 
we can take $k(C_n)=0$
and 
$f_n$ to be an arbitrary compressor
with $\ell(f_n(x^n))\geq 1$ for all $x^n$.
Then in either case,
$C_n^c=\{x^n:\ell(f_n(x^n))>k(C_n)\}$ 
and 
$\BBP(\ell(f_n(X^n))>k(C_n))=P_n(C_n^c)$,
so that,
\begin{align*}
\inf_{f_n,R} 
\left[\BBP(\ell(f_n(X^n)>nR)+\frac{2^{nR}}{|A|^n}\right]
&\leq
\inf_{f_n} 
\left[\BBP(\ell(f_n(X^n)>k(C_n))+\frac{2^{k(C_n)}}{|A|^n}\right]\\
&\leq
\inf_{C_n} 
\left[P_n(C_n^c)+\frac{|C_n|}{|A|^n}\right],
\end{align*}
which implies the upper bound in~(\ref{eq:Nfltonstar}).

For the lower bound we recall~\cite{kontoyiannis-verdu:14}
that, independently of the rate $R>0$,
the infimum over all compressors $f_n$ in the definition 
of $n^*(\Xp,\epsilon)$ is achieved by an $f_n^*$
that orders all strings $x^n$ in decreasing probability
(breaking ties arbitrarily), and sequentially
assigns to them binary codewords
of increasing length, lexicographically.
Then the $i$th most likely string $x^n$ has
$\ell(f_n^*(x^n))=\lfloor\log i\rfloor$.
For any $R>0$, letting $C_n=\{x^n:\ell(f_n^*(x^n))\leq nR\}$,
so that $|C_n|\leq 2^{\lfloor nR\rfloor +1}-1\leq2^{nR+1}$, we have:
\begin{align*}
\inf_{f_n,R} 
\left[\BBP(\ell(f_n(X^n)>nR)+\frac{2^{nR}}{|A|^n}\right]
&=
\inf_{R} 
\left[\BBP(\ell(f^*_n(X^n)>nR)+\frac{2^{nR}}{|A|^n}\right]\\
&\geq
\inf_{C_n} 
\left[P_n(C_n^c)+\frac{|C_n|}{2|A|^n}\right]\\
&\geq
\frac{1}{2}\inf_{C_n} 
\left[P_n(C_n^c)+\frac{|C_n|}{|A|^n}\right].
\end{align*}
This implies the lower bound in~(\ref{eq:Nfltonstar})
and completes the proof.
\qed

In order to examine the best achievable performance of 
prefix free codes, 
for any source $\Xp$ we
let $n^{\sf p}(\Xp,\epsilon)$ and $N^{\sf p}(\Xp,\epsilon)$
be defined exactly like $n^*(\Xp,\epsilon)$ and $N^*(\Xp,\epsilon)$,
in~(\ref{eq:nstar}) and~(\ref{eq:Nstar}), respectively,
but with the infima taken over all prefix-free compressors $f_n$.
It is not hard to show directly from the
definitions that the prefix-free requirement only induces
a minor degradation of compression performance:

\begin{theorem}[Variable-length vs.\ prefix-free sample complexity]
\label{thm:vltopf}
For any source $\Xp$ on $A$ and any $\epsilon\in(0,1)$,
\bqq
n^*(\Xp,\epsilon)\leq n^{\sf p}(\Xp,\epsilon)\leq n^*(\Xp,\epsilon/2),
\label{eq:nstarvsnp}
\eqq
while, for any $\epsilon\in(0,2)$,
\bqq
N^*(\Xp,\epsilon)\leq N^{\sf p}(\Xp,\epsilon)\leq N^*(\Xp,\epsilon/2).
\label{eq:NstarvsNp}
\eqq
\end{theorem}

\noindent
{\sc Proof.}
Recalling the definitions of the fundamental limits
$\epsilon^*(n,R)$ and $\epsilon^{\sf p}(n,R)$ from
the Introduction, we can express
\begin{align}
n^*(\Xp,\epsilon)
&=
	\inf\left\{n\geq 1\;:\;
	\inf_{R>0}\max\left\{\epsilon^*(n,R),\;\frac{2^{nR}}{|A|^n} \right\}
	\leq\epsilon\right\},
	\label{eq:nstaralt}\\
n^{\sf p}(\Xp,\epsilon)
&=
	\inf\left\{n\geq 1\;:\;
	\inf_{R>0}\max\left\{\epsilon^{\sf p}(n,R),\;\frac{2^{nR}}{|A|^n}
	\right\}\leq\epsilon\right\}.
	\label{eq:npalt}
\end{align}
These, together with the second inequality
in~(\ref{eq:epsilonstarp}) immediately imply 
that $n^*(\Xp,\epsilon)\leq n^{\sf p}(\Xp,\epsilon)$.
Similarly, using the first inequality in~(\ref{eq:epsilonstarp})
yields,
\begin{align*}
&
	\inf_{R>0}\max\left\{\epsilon^*(n,R),\;\frac{2^{nR}}{|A|^n} \right\}\\
&\geq
	\inf_{R>0}\max\left\{\epsilon^{\sf p}\Big(n,R+\frac{1}{n}\Big),
	\;\frac{2^{nR}}{|A|^n} \right\}\\
&\geq
	\inf_{R>0}\max\left\{\epsilon^{\sf p}(n,R),
	\;\frac{2^{nR-1}}{|A|^n} \right\}\\
&\geq
	\frac{1}{2}\inf_{R>0}\max\left\{\epsilon^{\sf p}(n,R),
	\;\frac{2^{nR}}{|A|^n} \right\}.
\end{align*}
This, combined with the expressions in~(\ref{eq:nstaralt})
and~(\ref{eq:npalt}), implies that 
$n^*(\Xp,\epsilon)\geq n^{\sf p}(\Xp,2\epsilon)$,
completing the proof of~(\ref{eq:nstarvsnp}).
The proof of~(\ref{eq:NstarvsNp}) is identical.
\qed

\subsection{Memoryless sources}
\label{s:VLdetail}

Using Theorems~\ref{thm:fltostar} and~\ref{thm:vltopf},
the sample complexity bounds for fixed-length compressors
established earlier in Theorem~\ref{thm:nfl}
easily translate to corresponding bounds for variable-length
and prefix-free compressors.
Theorem~\ref{thm:NstarIIDvl} essentially states that
$$n^*(\Xp,\epsilon)\;\mbox{and}\; n^{\sf p}(\Xp,\epsilon)
\;\mbox{are both}\;\;
\asymp\frac{\log(1/\epsilon)}{D_{1/2}(P\|U)}.$$
The bounds in Theorem~\ref{thm:NstarIIDvl} below
follow immediately from
Theorems~\ref{thm:fltostar},~\ref{thm:vltopf} and~\ref{thm:nfl}
through simple computations.

\begin{theorem}[Variable-length sample complexity of memoryless sources]
\label{thm:NstarIIDvl}
$\;$ Let $U$ denote the uniform p.m.f.\ on $A$. For any memoryless
source $\Xp$ with marginal p.m.f.\ 
$P\neq U$ on $A$ and any 
$\epsilon\in(0,1)$, the variable-length sample complexity
and the prefix-free sample complexity of $\Xp$ satisfy:
\begin{align}
\frac{\log(1/\epsilon)-3}{D_{1/2}(P\|U)}
&\leq n^*(\Xp,\epsilon)\leq
\frac{2\log(1/\epsilon)}{D_{1/2}(P\|U)}+1,
	\label{eq:nstarexplicit1}
	\\
\frac{\log(1/\epsilon)-3}{D_{1/2}(P\|U)}
&\leq n^{\sf p}(\Xp,\epsilon)\leq
\frac{2\log(1/\epsilon)+2}{D_{1/2}(P\|U)}+1.
	\nonumber
\end{align}
In particular, for $0<\epsilon<\min\{\frac{1}{8},\frac{1}{m}\},$
\begin{align}
\frac{\log(1/\epsilon)}{4D_{1/2}(P\|U)}
&\leq n^*(\Xp,\epsilon)\leq
\frac{3\log(1/\epsilon)}{D_{1/2}(P\|U)},
	\label{eq:nstarexplicit2}
	\\
\frac{\log(1/\epsilon)}{4D_{1/2}(P\|U)}
&\leq n^{\sf p}(\Xp,\epsilon)\leq
\frac{5\log(1/\epsilon)}{D_{1/2}(P\|U)}.
	\nonumber
\end{align}
\end{theorem}

We emphasize that Remarks~3--8 in Section~\ref{s:IID}
also apply
verbatim to the bounds in Theorem~\ref{thm:NstarIIDvl},
as well as to most of the 
sample complexity results
in subsequent sections.

\subsection{Non-asymptotic bounds versus Gaussian approximation}
\label{s:example}

A key difference between the sample complexity bounds
in Theorem~\ref{thm:NstarIIDvl} and the Gaussian approximation
\be
R^*(n,\epsilon)\approx
\widetilde{R}(n,\epsilon)
:= H(\Xp)+\sigma(\Xp)Q^{-1}(\epsilon)\frac{1}{\sqrt{n}}
-\frac{\log n}{2n},
\label{eq:strassen2}
\ee
suggested by the expansion we saw in~(\ref{eq:strassen})
in the Introduction, is that
the bounds in~(\ref{eq:nstarexplicit1}) are 
absolute whereas~(\ref{eq:strassen2}) is
only an asymptotic approximation.
More specifically, the Gaussian approximation
is expected to be more accurate in the
central limit theorem (CLT) regime, 
corresponding to moderate values of $\epsilon$,
and less accurate in the large deviations regime
when $\epsilon$ is close to zero;
see the relevant discussion in~\cite{theocharous:toappear}.
On the other
hand, the sample complexity bounds in~(\ref{eq:nstarexplicit1})
are uniformly valid over all $\epsilon\in(0,1)$.

For an explicit numerical comparison, we evaluate these 
two approaches in the case of a Bernoulli source $\Xp$, 
with parameter $p=0.11$ (the same example as
in~\cite{kontoyiannis-verdu:14}) and with $p=0.38$.
We plot the upper and lower
bounds to $n^*(\Xp,\epsilon)$ in~(\ref{eq:nstarexplicit1})
as a function of $\epsilon$,
as well as the Gaussian approximation 
$n_G(\epsilon)$ to $n^*(\Xp,\epsilon)$ obtained
as follows.
Inverting~(\ref{eq:strassen2})
by solving for $n$, we let $n_G(\epsilon)$ be
the smallest $n$ such that the 
rate $\widetilde{R}(n,\epsilon)$
satisfies the sample-complexity rate constraint
$2^{nR}/|A|^n\leq\epsilon$, i.e.,
such that
$$
\widetilde{R}(n,\epsilon)\leq\log|A|-\frac{1}{n}\log\frac{1}{\epsilon}.$$

\begin{figure}[!ht]
\centering
\includegraphics[width=4.4in]{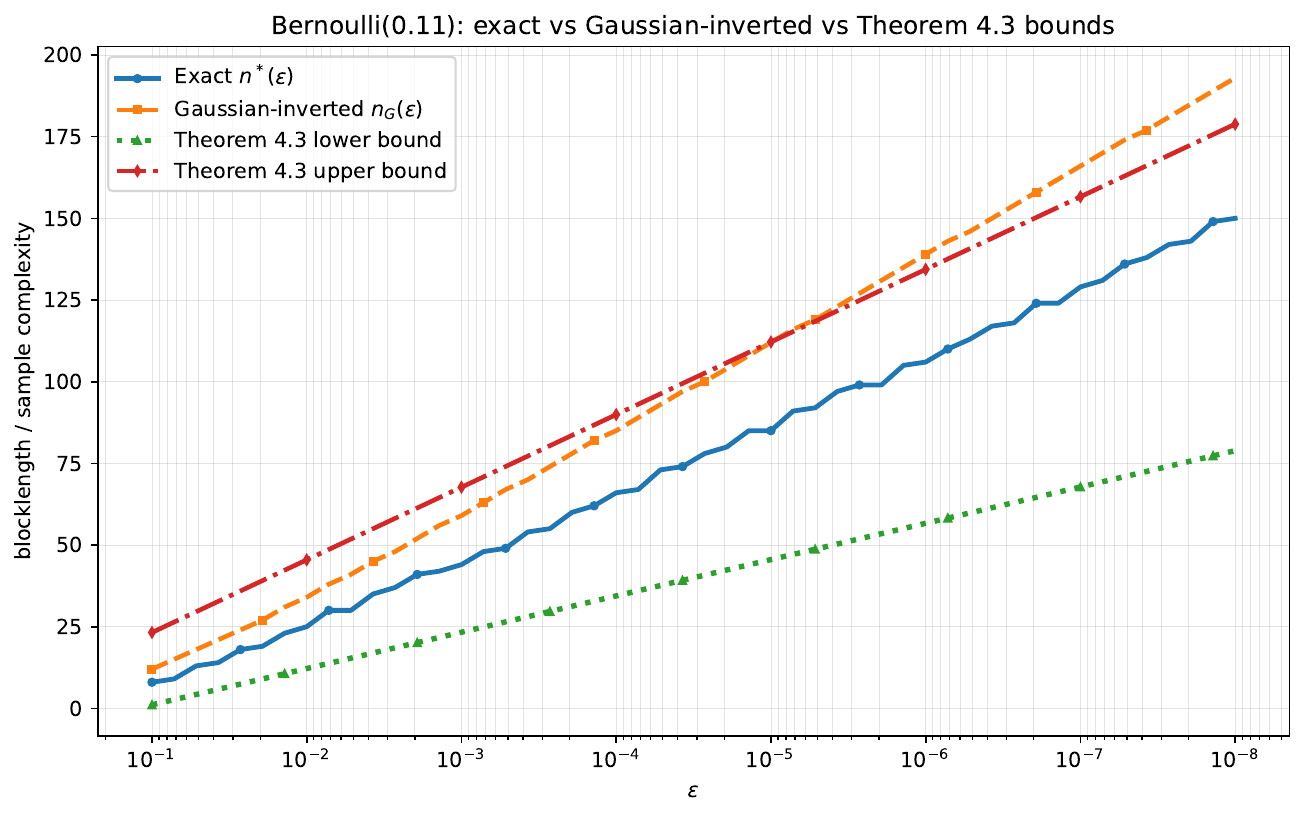}
\includegraphics[width=4.4in]{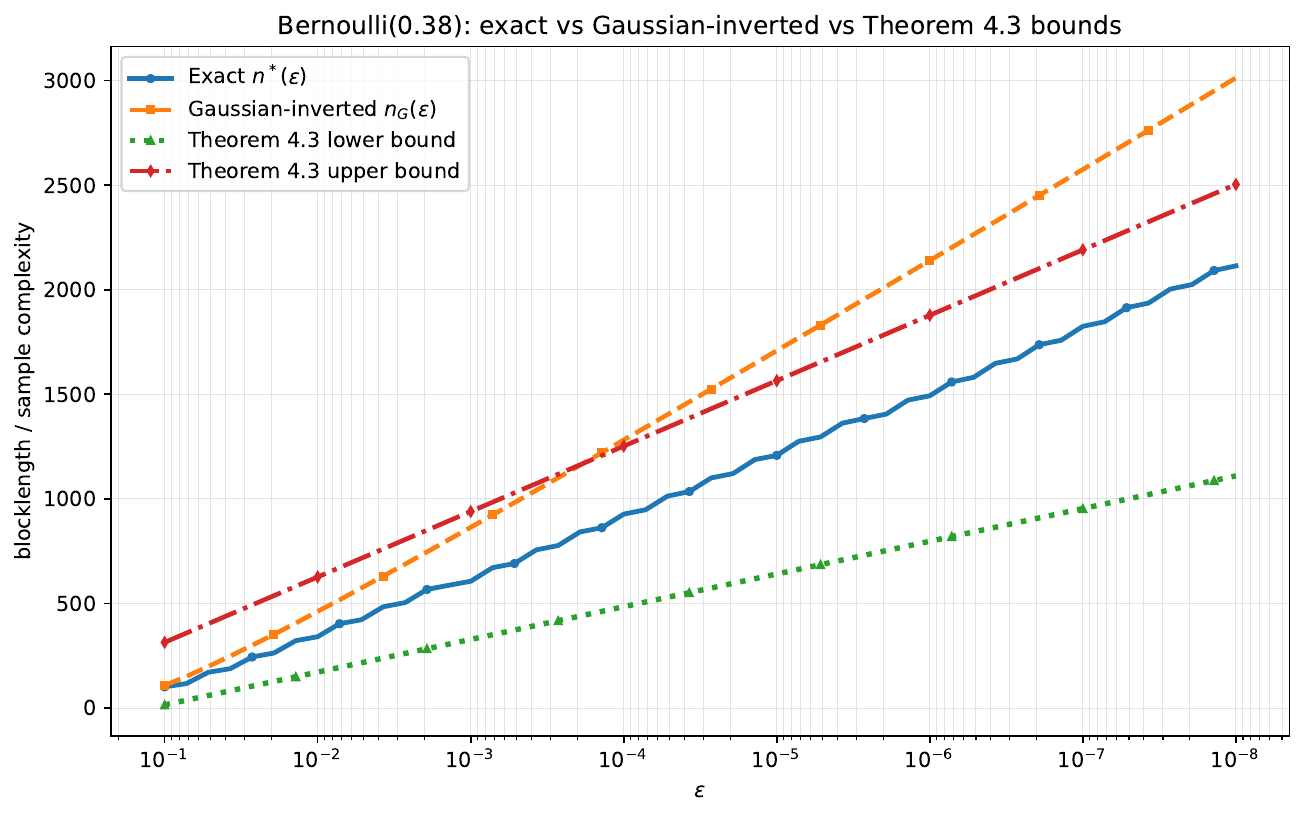}
\caption{Comparison of the sample complexity bounds 
on $n^*(\Xp,\epsilon)$
in Theorem~\ref{thm:NstarIIDvl},
with the Gaussian approximation estimate
$n_G(\epsilon)$ for $n^*(\Xp,\epsilon)$ 
based on the results in~\cite{kontoyiannis-verdu:14}.}
    \label{fig:example}
\end{figure}

The results, shown in Figure~\ref{fig:example}
along with the exact value of $n^*(\Xp,\epsilon)$,
clearly indicate that, as expected, 
$n^*(\Xp,\epsilon)$ is always between the upper
and lower bounds in 
Theorem~\ref{thm:NstarIIDvl},
and that the Gaussian approximation estimate $n_G(\epsilon)$
becomes significantly less accurate for smaller values of $\epsilon$.

\subsection{The actual compression rate}
\label{s:actual}

One of the central elements of the sample-complexity 
formulation of the data compression problem is that,
once a fixed bound $\epsilon>0$ is placed on the 
excess-rate probability, the best rate $R$ that
can be achieved is also automatically constrained 
to satisfy $2^{nR}/|A|^n\leq\epsilon$. Moreover, it
is easy to see that because of the monotonicity
of the excess-rate probability $\BBP\big(\ell(f_n(X^n))>nR\big)$
in $R$, the inner infimum over $R$ in the definition of the
fundamental limit $n^*(\Xp,\epsilon)$ in~\eqref{eq:nstar}
is always achieved with equality in the constraint,
i.e., at
$$R=\log|A|-\frac{1}{n}\log(1/\epsilon).$$
The traditional thinking in terms of asymptotically
large blocklengths would suggest that this rate
is typically close to $\log |A|$ so no compression
is achieved. But this is not at all the case:
The truly relevant rate in this formulation corresponds
to the case 
when the blocklength $n$ equals $n^*(\Xp,\epsilon)$.

\begin{figure}[!ht]
\centering
\includegraphics[width=3.9in]{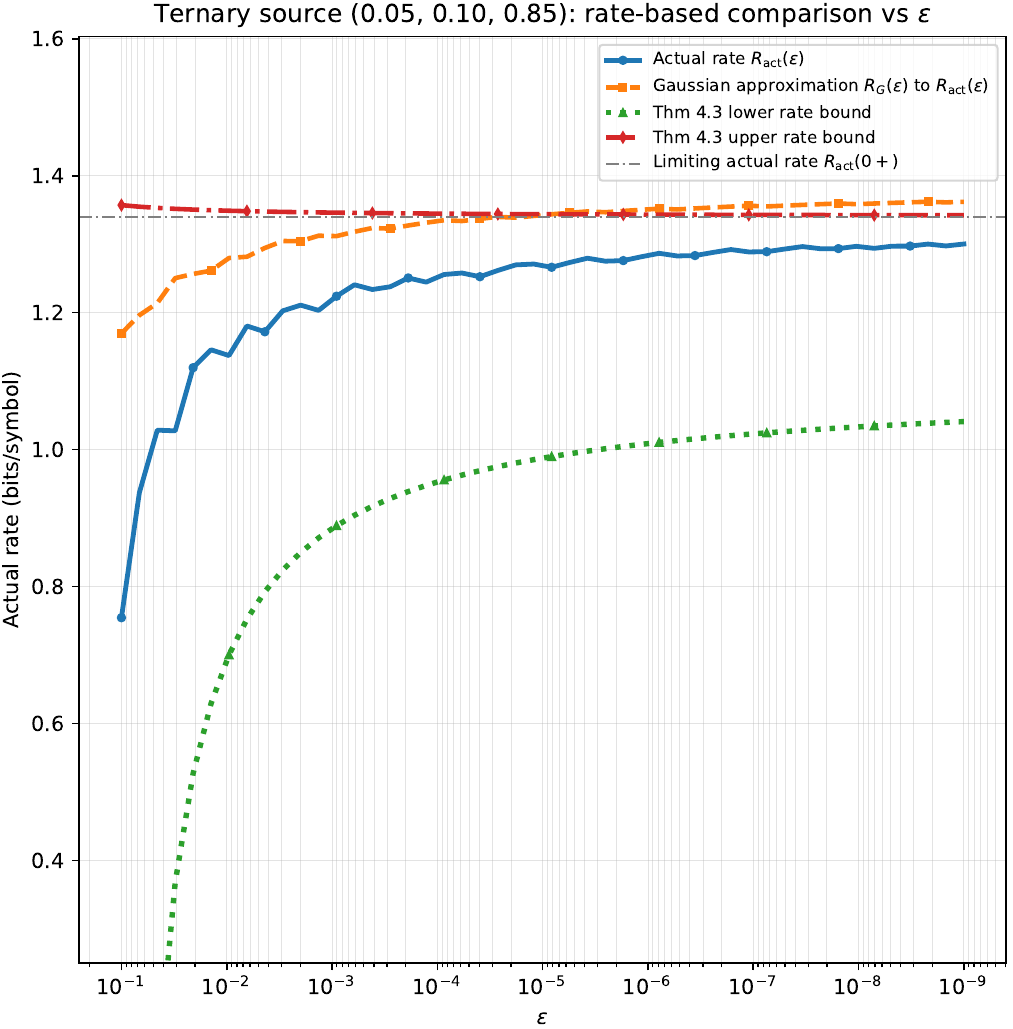}
\caption{Illustration of the
actual rate $R_{\rm act}(\epsilon)$ and its
limiting value $R_{\rm act}(0+)\approx 1.339$ bits/symbol
for a ternary memoryless
source with marginal p.m.f.\ $P=(0.05, 0.1, 0.85)$.
Also shown are the upper and lower bounds
on $R_{\rm act}(\epsilon)$ resulting
from the bounds 
in Theorem~\ref{thm:NstarIIDvl},
as well as the Gaussian approximation estimate
$R_G(\epsilon)$ to $R_{\rm act}(\epsilon)$
based on the results in~\cite{kontoyiannis-verdu:14}.}
    \label{fig:example2}
\end{figure}

We therefore define the {\em actual compression rate}
for a given $\epsilon>0$ as:
\be
R_{\rm act}(\epsilon)
:=\log|A|-\frac{1}{n^*(\Xp,\epsilon)}\log(1/\epsilon).
\label{eq:Ract}
\ee
Figure~\ref{fig:example2} illustrates 
$R_{\rm act}(\epsilon)$ as a function of $\epsilon$ for 
a ternary memoryless source $\Xp$
with marginal p.m.f.\ $P=(0.05,0.1,0.85)$ and with entropy
$H(P)\approx 0.6748$ bits.

The plot of $R_{\rm act}(\epsilon)$ in 
Figure~\ref{fig:example2} suggests
that, as $\epsilon\to 0$, the actual rate converges to a limiting
value $R_{\rm act}(0+)$. We next show that
this value can be computed in explicit form, and it admits a
natural representation in terms of the Chernoff information
$\clC(P,U)$ discussed earlier.

Note that our bounds only become looser as $\epsilon$ decreases,
because they scale as $\log (1/\epsilon)$. Looking,
as we do here, at the ratio of $n^\star$ and $\log(1/\epsilon)$ eliminates 
this problem. In this sense, the plots of Figure~\ref{fig:example2}
are more satisfying than those of Figure~\ref{fig:example}.

\begin{proposition}[Limiting actual rate]
\label{prop:actual}
If $P\neq U$ is a p.m.f.\ with full support on 
a finite alphabet $A$, then,
$$
R_{\rm act}(0+)=
\lim_{\epsilon\downarrow 0} R_{\rm act}(\epsilon)
=\log |A|-\clC(P,U),
$$
where $\clC(P,U)$ is the Chernoff information in~\eqref{eq:chernoff}.
\end{proposition}

Although the proof of Proposition~\ref{prop:actual},
given in Appendix~\ref{app:actual}, is rather technical,
the intuition behind it is simple. After relating
$n^*(\Xp,\epsilon)$ to the fixed-length
sample-complexity fundamental limit $N^{\sf fl}(\Xp,\epsilon)$,
we argue that the inner infimum in the definition
of $N^{\sf fl}(\Xp,\epsilon)$ in~\eqref{eq:Nfl}
is achieved when the two terms 
$\BBP(C_n^c)$ and $|C_n|/|A|^n$ decay at the same
exponential rate. Therefore, $R_{\rm act}(0+)$ is
the fixed point of the equation $E(R)=R$, where
$E(R)$ is the {\em reliability function}
of the error-exponents problem
for fixed-length data compression; cf.~\cite{csiszar:book}.

The bounds for $n^*(\Xp,\epsilon)$ 
in Theorem~\ref{thm:NstarIIDvl} can be substituted
in~\eqref{eq:Ract} to give corresponding upper 
and lower bounds for $R_{\rm act}(\epsilon)$. These
are shown in Figure~\ref{fig:example2} for the same
ternary source. Similarly, substituting
the Gaussian approximation
$n_G(\epsilon)$ for $n^*(\Xp,\epsilon)$ 
from the previous section
into~\eqref{eq:Ract}, gives a Gaussian 
approximation $R_G(\epsilon)$ to the
actual rate $R_{\rm act}(\epsilon)$, 
also shown in Figure~\ref{fig:example2}.

\subsection{More general target rates}
\label{s:asymmetric}

The key operational interpretation of the actual rate
$R_{\rm act}(\epsilon)$ in equation~\eqref{eq:Ract} 
above is as follows.
Once the excess-rate probability is constrained
to be below a given $\epsilon>0$, the sample-complexity 
formulation of the problem essentially {\em fixes}
the compression rate: It
{\em mandates} that optimal compression
at blocklength $n^*(\Xp,\epsilon)$ occurs
at rate $R_{\rm act}(\epsilon)$ bits/symbol.
Although this gives satisfying clarity from the theoretical 
point of view, the fact that it forces a specific rate
may be quite restrictive in practice.
However, there are several ways to overcome this limitation.

First, for moderate values of $\epsilon$, a simple approach
would be to proceed as in Remark~7 in Section~\ref{s:IID},
that is, examine the sample complexity when different
weights $\pi$ and $(1-\pi)$ are given to the two different 
objectives, i.e., to define $n^*$ in terms
of
$$\pi\BBP\big(\ell(f_n(X^n))>nR\big)\quad\mbox{and}\quad 
(1-\pi)\frac{2^{nR}}{|A|^n},$$
in place of simply
$\BBP\big(\ell(f_n(X^n))>nR\big)$ and $2^{nR}/|A|^n$,
for an appropriately chosen $\pi\in(0,1)$. Using
the recent hypothesis testing results in~\cite{pensia:24}
it would then be straightforward to obtain corresponding
upper and lower bounds to sample complexity in this
case, as well as to evaluate the resulting actual compression
rate.

For small values of $\epsilon$, the sample complexity 
(and hence the resulting blocklength) necessarily grows,
and the effects of the prior parameter $\pi$ 
get diminished. In this case, and in fact for the
entire range of $\epsilon\in(0,1)$, a more principled
approach that will lead to smaller actual rates,
closer to the entropy, is to define
for each $\beta>0$,
the {\em asymmetric variable-length sample complexity}
as:
$$n^*(\Xp,\epsilon;\beta)
=
	\inf\left\{n\geq 1\;:\;
	\inf_{f_n,R>0}\max\left\{
	\BBP(\ell(f_n(X^n))>nR),
	\frac{2^{n(R+\beta)}}{|A|^n}\right\}\leq\epsilon
	\right\}.
$$
In addition to the obvious fact that this formulation
leads to smaller ``actual rates'', it is an interesting
avenue for further work to develop upper and lower bounds 
on $n^*(\Xp,\epsilon;\beta)$ analogous to those in 
Theorem~\ref{thm:NstarIIDvl}, and to study the behaviour
of the {\em asymmetric actual rate},
$$R_{\rm act}(\epsilon;\beta):=\log|A|-\beta-\frac{1}{n^*(\Xp,\epsilon;\beta)}
\log(1/\epsilon),$$
as $\epsilon\downarrow 0$.
We will not pursue this direction further here,
except to note the following simple upper bound
on $n^*(\Xp,\epsilon;\beta)$, in the spirit
of the corresponding upper bound
on $n^*(\Xp,\epsilon)$ in 
Theorem~\ref{thm:NstarIIDvl}: 

\begin{proposition}[Asymmetric variable-length
sample complexity upper bound] \
Let $\Xp$ be
a memoryless source with marginal p.m.f.\ $P$
on $A$. Then,
for any $0<\beta<D(P\|U)$ and 
for any $\epsilon\in(0,1)$, we have,
$$n^*(\Xp,\epsilon;\beta)
\leq\frac{\log(1/\epsilon)}{
(1-\alpha)[D_{\alpha}(P\|U)-\beta)]^+
}+1,$$
for any $\alpha\in(0,1)$.
\end{proposition}

\noindent
{\sc Proof.}
Fix $0<\beta<D(P\|U)$. Without loss of generality, 
we can consider only those $\alpha\in(0,1)$ for which
$D_\alpha(P\|U)-\beta$ is strictly positive. Choose and
fix any such $\alpha$.

First we examine the minimum
$$\epsilon(n)=\min_{C_n\subset A^n}\big[P^n(C_n^c)+2^{n\beta}U^n(C_n)\big].$$
By the definition of $\epsilon(n)$ and using the elementary
inequality $\min\{x,y\}\leq x^{\alpha}y^{1-\alpha}$, $x,y\geq 0$, we have
\begin{align*}
\epsilon(n)
&=
	\sum_{x^n\in A^n}
	\min\left\{P^n(x^n),
	2^{n\beta}U^n(x^n)\right\}
	\\
&\leq
	2^{n\beta(1-\alpha)}\sum_{x^n\in A^n}
	\big[P^n(x^n)]^{\alpha}
	\big[U^n(x^n)\big]^{1-\alpha}
	\\
&=
	2^{n\beta(1-\alpha)}2^{-(1-\alpha)D_\alpha(P^n\|U^n)}.
\end{align*}
Using the tensorization property of R\'{e}nyi divergence in
Proposition~\ref{prop:tensorize}, gives
$$\epsilon(n)\leq 2^{-n(1-\alpha)[D_{\alpha}(P\|U)-\beta]}.$$
Therefore, $\epsilon(n)\leq \epsilon$ 
for all
$$n\geq \frac{\log(1/\epsilon)}{(1-\alpha)[D_\alpha(P\|U)-\beta]},$$
which means that there is a $C_n\subset A^n$ with 
both 
\be
P^n(C_n^c)\leq\epsilon\quad\mbox{and}\quad 2^{n\beta}\frac{|C_n|}{|A|^n}
\leq \epsilon.
\label{eq:asymmetric}
\ee
Let $R=\frac{1}{n}\log |C_n|$, and define a compressor
$f_n$ that maps all $x^n\in C_n$ to binary strings of length
no greater than $\log|C_n|=nR$, and all other $x^n$ to arbitrary
binary strings. Noting that in this case
$\BBP(\ell(f_n(X^n))>nR)\leq P^n(C_n^c),$
the definition of $n^*(\Xp,\epsilon;\beta)$ and the
bounds~\eqref{eq:asymmetric} give the claimed upper bound
on $n^*(\Xp,\epsilon;\beta)$.
\qed

We close this section with 
two simple observations.
First, by the definition of the R\'{e}nyi divergence and the 
assumption that $0<\beta<D(P\|U)$, there is always an interval
of values $\alpha$ in $(0,1)$ for which $D_\alpha(P\|U)-\beta$
is strictly positive, so the upper bound in the proposition is
not vacuous.
Second, since by definition we always have that
$n^*(\Xp,\epsilon;\beta)\geq n^*(Xp,\epsilon)$, 
by Theorem~\ref{thm:NstarIIDvl}
we also
have the elementary lower bound
$$n^*(\Xp,\epsilon;\beta)\geq\frac{\log(1/\epsilon)-3}{D_{1/2}(P\|U)}.$$

\section{Markov sources}
\label{s:markov}

In this section we examine the sample complexity
of compressing Markov sources. Since the corresponding
literature in hypothesis testing is much more limited
than in the i.i.d.\ case,
and since the problem itself is intrinsically harder,
more effort is required to obtain useful bounds
on $n^*(\Xp,\epsilon)$ when $\Xp$ is a Markov chain.

In Section~\ref{s:generalmarkov} we derive general
bounds on $n^*(\Xp,\epsilon)$ for any irreducible
Markov source $\Xp$; these 
depend not just on the R\'{e}nyi divergence rate
of the source, 
but also on its initial distribution and on the
right Perron eigenvector of a matrix associated
with its transition matrix.
More explicit bounds that, like those obtained
for memoryless sources,
only depend on R\'{e}nyi divergence are established
for the special case of {\em symmetric} Markov chains
in Section~\ref{s:symmetric}.

Since, as we saw in Section~\ref{s:VLgeneral},
it is easy to translate results between
$n^*(\Xp,\epsilon)$, $n^{\sf fl}(\Xp,\epsilon)$ and
$n^{\sf p}(\Xp,\epsilon)$, in this section we only
consider the variable-length
sample complexity $n^*(\Xp,\epsilon)$ of Markov chains
$\Xp$.

\subsection{R\'{e}nyi divergence rate and irreducible Markov sources}
\label{s:generalmarkov}

Let $\Xp=\{X_n\;;\;n\geq 1\}$ be a Markov chain on $A=\{a_1,\ldots,a_m\}$
with initial distribution $\mu$
and transition matrix $\bfP=(p_{ij})_{1\leq i,j\leq m}$,
so that
$\BBP(X_1=a_i)=\mu(a_i)$ 
and 
$\BBP(X_{n+1}=a_j|X_n=a_i)=p_{ij}$, 
for $a_i,a_j\in A$ and $n\geq 1$.

The following notation will be useful throughout this section.
For any two column vectors $u=(u_1,\ldots,u_m)^\transposed,
v=(v_1,\ldots,v_m)^\transposed\in\RL^m$, we write
$u\odot v$ for their element-wise product, so that
$u\odot v\in\RL^m$ with $(u\odot v)_i=u_iv_i$, $1\leq i\leq m$.
Similarly, $\bfR\odot \bfS$ denotes the element-wise product of two
$m\times m$ matrices $\bfR,\bfS$. And for a nonnegative vector $v$ 
or a nonnegative matrix $\bfR$,
we write $v\subsqrt$ and $\bfR\subsqrt$ for the corresponding
vector or matrix with elements given by the square-root
of its original elements; for example,
$(v\subsqrt)_i=\sqrt{v_i}$.

A nonnegative $m\times m$ matrix $\bfR$ is {\em irreducible}
if for any pair of indices $1\leq i,j\leq m$ there is
an integer $k$ such that $(\bfR^k)_{ij}>0$.
The Perron-Frobenius theorem~\cite{seneta:81}
states that any nonnegative irreducible $m\times m$
matrix $\bfR$ has a real and positive eigenvalue 
$\lambda=\lambda(\bfR)$ of maximal modulus, 
whose associated left and right
eigenvectors $u$ and $v$ have strictly positive elements.
We call $\lambda$ and $u,v$ the {\em Perron eigenvalue}
and the {\em Perron eigenvectors} of $\bfR$, respectively.

In the bounds derived in this section,
the R\'{e}nyi divergence $D_{1/2}(P\|U)$
is replaced by the R\'{e}nyi divergence
{\em rate} $D_{1/2}(\Xp\|\Up)$, between
a Markov source $\Xp$ and the i.i.d.\ uniform
source $\Up$. The following results
for $D_{1/2}(P_n\|Q_n)$ and $D_{1/2}(\Xp\|\Yp)$
between two Markov sources $\Xp$ and $\Yp$ were
derived in~\cite{rached:01};
see also~\cite{daskalakis:18,cherapanamjeri:19} 
for related computations.
The expression~(\ref{eq:DPnQn}) follows by
direct calculation and induction on $n$,
and~(\ref{eq:DXYrate}) follows from~(\ref{eq:DPnQn})
combined with the Perron-Frobenius theorem.

\begin{proposition}[R\'{e}nyi divergence of Markov chains]
\label{prop:renyidiv}
Suppose $\Xp,\Yp$ are Markov chains on the same
finite alphabet $A$, with initial distributions
$\mu,\nu$ and transition matrices $\bfP,\bfQ$, respectively.
For each $n\geq 1$, letting 
$P_n,Q_n$ denote the $n$-dimensional marginal distributions
of $X^n$ and $Y^n$, respectively, we have
\bqq
D_{1/2}(P_n \| Q_n) 
= -2\log \left( [\mu \odot \nu]\subsqrt^\transposed  
[\bfP \odot \bfQ]\subsqrt^{n-1}  \bfone \right),
\label{eq:DPnQn}
\eqq
where we view $\mu$ and $\nu$ as column vectors in $\RL^m$,
and $\bfone\in\RL^m$ denotes the all-1 column vector.

Moreover, if the matrix $[\bfP\odot\bfQ]$ is irreducible,
then:
\bqq
D_{1/2}(\Xp\|\Yp)=\lim_{n\to\infty}\frac{1}{n}D_{1/2}(P_n\|Q_n)
=-2\log\lambda\big([\bfP\odot\bfQ]\subsqrt\big).
\label{eq:DXYrate}
\eqq
\end{proposition}

We are now in a position to state our first result
on the sample complexity of Markov sources.
Its proof is based on computations similar
to those carried out in~\cite{rached:01}.

\medskip

\begin{theorem}[Sample complexity of irreducible Markov sources]
\label{thm:irreducible}
$\;$ Let $\Up$ denote the i.i.d.\ uniform process on an 
alphabet $A$ of size $|A|=m$.
Suppose 
$\Xp\neq\Up$
is a Markov source with initial distribution $\mu$
and irreducible transition matrix $\bfP$,
and write $w$ for the vector $\mu\subsqrt$.
Then the variable-length sample complexity of $\Xp$ satisfies,
for each $\epsilon\in(0,1)$,
$$\frac{ \log(1/\epsilon) + 2\log 
\Big( \frac{w^\transposed v}{\sqrt{m}\bar{v}} \Big) - 4}
{D_{1/2}(\Xp\|\Up)}+1 
\leq n^*(\Xp,\epsilon) 
\leq \frac{ 2\log ( 1/\epsilon) 
+ 2\log \Big( \frac{w^\transposed v}
{\sqrt{m}\underline{v}} \Big)}
{D_{1/2}(\Xp\|\Up)}+2,$$
where $v$ is the unit-norm right Perron eigenvector of 
the matrix $[\frac{1}{m}\bfP]\subsqrt$,
$\bar{v}=\max_{1\leq i\leq m} v_i$, 
and $\underline{v}=\min_{1\leq i\leq m}v_i$.
\end{theorem}

\noindent
{\sc Proof.}
We can view $\Up$ as a Markov chain with initial distribution
$\nu=U$ and transition matrix $\bfQ=(q_{ij})$ where $q_{ij}=1/m$
for all $i,j$. 
Since $\bfP$ is irreducible, so is
the matrix $[\bfP\odot\bfQ]\subsqrt=[\frac{1}{m}\bfP]\subsqrt$, and by 
the Perron-Frobenius theorem it has
a positive right eigenvector $v$ with $\|v\|_2 = 1$,
corresponding to its Perron eigenvalue
$\lambda=\lambda([\bfP\odot\bfQ]\subsqrt)$.
Therefore, for all $n \geq 1$,
$$[\bfP \odot \bfQ]\subsqrt^{n-1}v = \lambda^{n-1}v,$$
and since $[\bfP \odot \bfQ]\subsqrt$ is nonnegative, we have,
$$\underline{v} [\bfP \odot\bfQ]\subsqrt^{n-1} \bfone
\leq [\bfP \odot \bfQ]\subsqrt^{n-1}v 
\leq \bar{v}[\bfP \odot \bfQ]\subsqrt^{n-1}  \bfone.$$
Combining the last two expressions, rearranging,
and multiplying by $[\mu\odot \nu]\subsqrt^\transpose$
throughout, yields,
$$\frac{\lambda^{n-1} [\mu \odot \nu]\subsqrt^\transposed v}{\bar{v}} 
\leq [\mu \odot \nu]\subsqrt^\transposed  [\bfP \odot \bfQ]\subsqrt^{n-1} 
 \bfone
\leq \frac{\lambda^{n-1} [\mu\odot \nu]\subsqrt^\transposed v}
{\underline{v}}.$$
Taking logarithms and recalling that $\nu=U=\frac{1}{m}\bfone$,
$Q_n=U^n$ 
and $w=\mu\subsqrt$, gives:
\begin{align}
-2(n-1)\log \lambda  - 
2\log \Big( \frac{w^\transposed v}{\sqrt{m}\underline{v}} \Big) 
& \leq D_{1/2}(P_n\|U^n) 
	\label{eq:PFbound1}\\ 
& \leq -2(n-1)\log \lambda  - 2\log
\Big( \frac{w^\transposed v}{\sqrt{m}\bar{v}} \Big).
\label{eq:PFbound2}
\end{align}

Now, as in the proof of Theorem~\ref{thm:nfl},
let $\epsilon(n) = 1-\|P_n-U^n\|_{\rm TV}$.
By Proposition~\ref{prop:DTV},
\bqq
2^{-D_{1/2}(P_n\|U^n)-1}\leq\epsilon(n)
\leq 2^{-\frac{1}{2}D_{1/2}(P_n\|U^n)}.
\label{eq:prelim}
\eqq
Recalling from Proposition~\ref{prop:renyidiv}
that $D_{1/2}(\Xp\|\Up)=-\log\lambda$ 
and using our earlier bounds 
in~(\ref{eq:PFbound1})
and~(\ref{eq:PFbound2}) on $D_{1/2}(P_n\|Q_n)$,
the bounds~(\ref{eq:prelim}) become
$$2^{-(n-1)D_{1/2}(\Xps\|\Ups)
+ 2\log\big( \frac{w^\transposed v}{\sqrt{m}\bar{v}} \big)-1} 
\leq \epsilon(n)
\leq 2^{-\frac{1}{2}(n-1){D_{1/2}(\Xps\|\Ups)}
+ \log \big( \frac{w^\transposed v}{\sqrt{m}\underline{v}} \big)}.$$
Le Cam's lemma then implies that 
$$\frac{ \log(1/\epsilon) + 2\log 
\Big( \frac{w^\transposed v}{\sqrt{m}\bar{v}} \Big) - 2}
{D_{1/2}(\Xp\|\Up)}+1 
\leq N^{\sf fl}(\Xp,\epsilon) 
\leq \frac{ 2\log ( 1/\epsilon) 
+ 2\log \Big( \frac{w^\transposed v}
{\sqrt{m}\underline{v}} \Big)}
{D_{1/2}(\Xp\|\Up)}+2.$$
Theorem~\ref{thm:fltostar} implies that the
same result holds for $N^*(\Xp,\epsilon)$
in place of $N^{\sf fl}(\Xp,\epsilon)$,
and the observation~(\ref{eq:Nstarnstar})
gives the claimed result for $n^*(\Xp,\epsilon)$.
\qed

As irreducible Markov chains include all ergodic chains,
Theorem~\ref{thm:irreducible} is about as general as 
one might hope for. But the bounds themselves are not as satisfying
as those obtained earlier for memoryless sources,
in that they depend on finer properties of the source
than just its R\'{e}nyi divergence rate, namely,
on its initial distribution $\mu$ and the right Perron
eigenvector of the matrix $[\frac{1}{m}\bfP]\subsqrt$.
In fact, it is easy to construct examples
with fixed R\'{e}nyi divergence rates $D_{1/2}(\Xp\|\Up)$ 
but such that $\bar{v}$ is close to~1 while $\underline{v}$ 
is close to zero. In that case the
second terms in the upper and lower bounds are very different.
In the following section we derive
simpler, much more explicit and tighter bounds for an
important special class of Markov sources.

\subsection{Symmetric Markov sources}
\label{s:symmetric}

A Markov chain $\Xp$ is {\em symmetric} if its transition matrix
is symmetric, $\bfP^\transposed=\bfP$, in which case the uniform
distribution is invariant for $\Xp$.
The proof of the following theorem follows closely along 
the lines of identity-testing arguments in~\cite{daskalakis:18}.

\begin{theorem}[Sample complexity of symmetric Markov sources]
\label{samplecomplexitysymmetric}
$\;$ Let $\Up$ denote the i.i.d.\ uniform process on an alphabet
$A$ of size $|A|=m$. If 
$\Xp\neq \Up$
is a symmetric, irreducible Markov source 
with initial distribution $\mu=U$,
then, for each $\epsilon\in(0,1)$:
$$\frac{\log(1/\epsilon) - 2\log m-4}
{D_{1/2}(\Xp\|\Up)} 
\leq n^*(\Xp,\epsilon) 
\leq \frac{2\log(1/\epsilon)+2\log m }
{D_{1/2}(\Xp\|\Up)}+3.$$
\end{theorem}

\noindent
{\sc Proof.}
Let $\bfP$ denote the transition matrix of $\Xp$.
As in the previous proof,
we view $\Up$ as a Markov chain with initial distribution
$\nu=U$ and transition matrix $\bfQ=(q_{ij})$ where $q_{ij}=1/m$
for all $i,j$. 
Since $\bfP$ is symmetric, so is
$[\bfP \odot \bfQ]\subsqrt = [\frac{1}{m}\bfP]\subsqrt$,
and by the spectral theorem it has
$m$ real eigenvalues  $\lambda_1\geq \cdots\geq \lambda_m$,
with corresponding  eigenvectors
$v(1),\ldots,v(m)$ that form 
an orthonormal basis of $\RL^m$,
and diagonalise $[\bfP\odot\bfQ]\subsqrt$ as:
$$[\bfP \odot \bfQ]\subsqrt = \sum_{i=1}^m \lambda_i v(i)v(i)^\transposed.$$
By the Perron-Frobenius theorem, $|\lambda_1|\geq|\lambda_i|$ 
for all $i$, and $v(1)$ can be chosen to have strictly positive entries.

Define the constants,
$$\tau_i = \Big[\frac{1}{m}\mu\Big]\subsqrt^\transposed v(i)v(i)^\transposed 
\bfone = \frac{1}{m} ( v(i)^\transposed \bfone )^2,
\quad 1\leq i\leq m.$$ 
Using Proposition~\ref{prop:renyidiv}, for each $n\geq 1$
the R\'{e}nyi divergence between $P_n$ and $U^n$ can then
be written,
\bqq
D_{1/2}( P_n \| U^n) = -2 \log\left( \sum_{i=1}^m \lambda_i^{n-1} \tau_i 
\right),
\label{eq:symmetricD}
\eqq
and the corresponding 
R\'{e}nyi divergence rate is 
\bqq
D_{1/2}(\Xp\|\Up)=-2\log\lambda_1.
\label{eq:symmetricDrate}
\eqq
Since $\{v(1),\ldots,v(m)\}$ form an orthonormal basis
we have $\tau_i\leq 1$ for all $i$,
and since the entries of $v(1)$ are positive we have
$\tau_1 \geq \frac{1}{m}$.
If $n$ is odd, then  $\lambda_i^{n-1} \geq 0$ for all $n\geq 1$,
$1\leq i\leq m$, and we can obtain the bounds:
$$\frac{\lambda_1^{n-1}}{m} \leq \lambda_1^{n-1} \tau_1 
\leq \sum_{i=1}^m \lambda_i^{n-1} \tau_i 
\leq \sum_{i=1}^m \lambda_i^{n-1} 
\leq m \lambda_1^{n-1}.$$
Combining these with~(\ref{eq:symmetricD}) and~(\ref{eq:symmetricDrate}),
yields, for $n$ odd, 
\bqq
(n-1)D_{1/2}(\Xp\|\Up) - 2\log  m 
\leq D_{1/2}( P_n \| U^n) \leq 
(n-1)D_{1/2}(\Xp\|\Up) + 2\log  m.
\label{eq:symmpre}
\eqq
But also, 
the sequence $D_{1/2}(P_n \| U^n)=n\log|A|-H_{1/2}(P_n)$ is 
easily seen~\cite{vanerven:14} to be nondecreasing 
in $n$. Therefore, for all $n$ we have
$$(n-2)D_{1/2}(\Xp\|\Up) - 2\log  m 
\leq D_{1/2}( P_n \| U^n) 
\leq nD_{1/2}(\Xp\|\Up) + 2\log m.$$

Finally,
as in the proof of Theorem~\ref{thm:nfl},
let $\epsilon(n) = 1-\|P_n-U^n\|_{\rm TV}$
so that, by 
Proposition~\ref{prop:DTV},
$$
2^{-D_{1/2}(P_n\|U^n)-1}\leq\epsilon(n)
\leq 2^{-\frac{1}{2}D_{1/2}(P_n\|U^n)}
$$
and by~(\ref{eq:symmpre}),
$$2^{-nD_{1/2}(\Xps\|\Ups) - 2\log  m - 1} 
\leq \epsilon(n)
\leq 2^{-\frac{(n-2)}{2}D_{1/2}(\Xps\|\Ups)+ \log m}.$$
Le Cam's lemma then implies that 
$$\frac{\log(1/\epsilon) - 2\log m-2}
{D_{1/2}(\Xp\|\Up)} 
\leq N^{\sf fl}(\Xp,\epsilon) 
\leq \frac{2\log(1/\epsilon)+2\log m }
{D_{1/2}(\Xp\|\Up)}+3,$$
and combining this with Theorem~\ref{thm:fltostar} 
and~(\ref{eq:Nstarnstar})
gives the claimed result for $n^*(\Xp,\epsilon)$.
\qed

As in the case of memoryless sources,
the sample complexity of compressing 
a Markov source
is not determined by its entropy rate,
but rather by the R\'{e}nyi divergence
rate $D_{1/2}(\Xp\|\Up)$ or, equivalently,
by the source's R\'{e}nyi entropy rate
$H_{1/2}(\Xp)=\lim_n\frac{1}{n}H_{1/2}(P_n)$,
since we always have $D_{1/2}(\Xp\|\Up)=
\log m -H_{1/2}(\Xp)$.

\section{Universal compression}
\label{s:universal}

\subsection{Arbitrary sources}
\label{s:universalG}

In this section we consider the sample complexity
of {\em universal} compression of an arbitrary
collection $\clS$ of sources $\Xp$ with
values in a given 
finite alphabet $A$.
Specifically, we ask for the shortest blocklength $n$
for which there is a variable-length compressor $f_n$
that achieves excess-rate probability no greater than
$\epsilon$ for {\em every} source in $\clS$, at some rate
$R$ such that $2^{nR}/|A|^n\leq \epsilon$:
For every $\epsilon\in(0,1)$, we define 
the {\em universal fixed-length 
sample complexity} of the family $\clS$ as
$$n^{\sf fl}(\clS,\epsilon)
=\inf\left\{n\geq 1\;:\;
\inf_{C_n\subset A^n}
\max\left\{\sup_{\Xps\in\clS}
\BBP\big(X^n\in C_n^c\big),\;\frac{|C_n|}{|A|^n}
\right\}
\leq\epsilon\right\},
$$
and the {\em universal variable-length 
sample complexity} of the family $\clS$ as
$$n^*(\clS,\epsilon)
=\inf\left\{n\geq 1\;:\;
\inf_{f_n,R>0}\max\left\{\sup_{\Xps\in\clS}
\BBP\big(\ell(f_n(X^n))>nR\big),\;\frac{2^{nR}}{|A|^n}
\right\}\leq\epsilon\right\}.
$$
The corresponding fundamental limits
$N^*(\clS,\epsilon)$,
$N^{\sf fl}(\clS,\epsilon)$,
$n^{\sf p}(\clS,\epsilon)$, and
$N^{\sf p}(\clS,\epsilon)$
are defined in the obvious way,
in analogy to their counterparts
in the case of a single source $\Xp$.
From the definitions, we immediately have,
as in the case of a known source that:
\begin{align}
N^{\sf fl}(\clS,2\epsilon)
&\leq 
	n^{\sf fl}(\clS,\epsilon)
	\leq N^{\sf fl}(\clS,\epsilon),
	\nonumber\\
N^*(\clS,2\epsilon)
&\leq 
	n^*(\clS,\epsilon)
	\leq N^*(\clS,\epsilon),
	\nonumber\\
N^{\sf p}(\clS,2\epsilon)
&\leq 
	n^{\sf p}(\clS,\epsilon)
	\leq N^{\sf p}(\clS,\epsilon).
	\nonumber
\end{align}
The simple bounds in
Theorems~\ref{thm:fltostar} and~\ref{thm:vltopf}
also extend to the case of universal compression.
In order to state and prove them, we find it useful
to define universal versions of the fundamental
limits
$\epsilon^*(n,R)$ and $\epsilon^{\sf p}(n,R)$ defined
in Section~\ref{s:generalf}.

Let $\clS$ be an arbitrary family of sources
on $A$. We define the 
best universally achievable excess-rate
probability on $\clS$ at a given rate $R>0$
and blocklength $n$ as
\bqq
\epsilon^*(\clS,n,R)=\inf_{f_n}
\sup_{\Xps\in\clS}\BBP\big(\ell(f_n(X^n)) >nR\big).
\label{eq:epsilonstarU}
\eqq
In the case of prefix-free codes, 
$\epsilon^{\sf p}(\clS,n,R)$ is
similarly defined, with the infimum
in~(\ref{eq:epsilonstarU}) taken
over all prefix-free compressors $f_n$.

The results in the following theorem are proved
based on combinatorial arguments similar to those 
in the proof of Theorem~\ref{thm:fltostar},
and part~$(iii)$ follows 
from~\cite[Theorem~1]{kontoyiannis-verdu:14}.

\begin{theorem}[Fundamental limits for universal compression]
\label{thm:universalFL}
Let $A$ be a finite alphabet and let $\clS$ be an arbitrary class 
of sources on $A$.
\begin{itemize}
\item[$(i)$]
For any $\epsilon\in(0,1)$ we have,
\begin{align}
n^{\sf fl}(\clS,2\epsilon) 
&\leq 
	n^*(\clS,\epsilon)
	\leq n^{\sf fl}(\clS,\epsilon),
	\label{eq:uni1}\\
N^{\sf fl}(\clS,2\epsilon) 
&\leq 
	N^*(\clS,\epsilon)
	\leq N^{\sf fl}(\clS,\epsilon),
	\label{eq:uni2}
\end{align}
with the understanding that 
$n^{\sf fl}(\clS,\epsilon)=N^{\sf fl}(\clS,\epsilon)=1$
for $\epsilon\geq 1$.
\item[$(ii)$]
For any $R>0$ and all $n\geq 1$:
\bqq
\epsilon^{\sf p}\Big(\clS,n,R+\frac{1}{n}\Big)
\leq\epsilon^*(\clS,n,R)
\leq\epsilon^{\sf p}(\clS,n,R).
\label{eq:epsilonstarpU}
\eqq
\item[$(iii)$]
For any $\epsilon\in(0,2)$ we have,
\begin{align}
n^*(\clS,\epsilon)
&\leq 
	n^{\sf p}(\clS,\epsilon)
	\leq n^*(\clS,\epsilon/2),
	\label{eq:uni3}\\
N^*(\clS,\epsilon)
&\leq 
	N^{\sf p}(\clS,\epsilon)\leq N^*(\clS,\epsilon/2),
	\label{eq:uni4}
\end{align}
with the understanding that
$n^*(\clS,\epsilon)=n^{\sf p}(\clS,\epsilon)=
N^*(\clS,\epsilon)=N^{\sf p}(\clS,\epsilon)=1$
for $\epsilon\geq 1$.
\end{itemize}
\end{theorem}

\noindent
{\sc Proof.}
$(i)$:
The proof of~(\ref{eq:uni1}) is very similar to the 
proof of Theorem~\ref{thm:fltostar}.
For the upper bound, given any 
$C_n\subset A^n$, there is a 
compressor $f_n$ that maps the elements of $C_n$
to binary strings of length no larger than 
$k(C_n)=\lfloor\log|C_n|\rfloor$,
and all elements of $C_n^c$ to arbitrary
binary strings of length at least $k(C_n)+1$.
Then,
with the same caveat about the case
where $C_n$ is empty as in the proof of Theorem~\ref{thm:fltostar},
$\BBP(\ell(f_n(X^n))>k(C_n))
\leq \BBP(X^n\in C^c_n))$
for any source $\Xp\in\clS$, hence,
\begin{align*}
\inf_{R>0}
\max
	\left\{\sup_{\Xps\in\clS}\BBP(\ell(f_n(X^n)>nR),
	\;\frac{2^{nR}}{|A|^n}\right\}
&\leq
\max
	\left\{\sup_{\Xps\in\clS}\BBP(\ell(f_n(X^n)>k(C_n)),
	\;\frac{2^{k(C_n)}}{|A|^n}\right\}\\
&\leq
	\max
	\left\{\sup_{\Xps\in\clS}\BBP(X^n\in C_n^c),\;
	\frac{|C_n|}{|A|^n}\right\}.
\end{align*}
Taking the infimum over all $f_n$ on the left-hand side
and over all $C_n$ on the right hand side,
and using 
the definitions
of $n^{\sf fl}(\clS,\epsilon)$ and 
$n^*(\clS,\epsilon)$, gives the 
upper bound in~(\ref{eq:uni1}).

For the lower bound, 
consider any $R>0$ and any compressor
$f_n$.
Let $C_n$ consist of all strings $x^n$ such that
$\ell(f_n(x^n))\leq nR$, so that 
$|C_n|\leq 2^{nR+1}$.
Then for any $\Xp\in\clS$ we have $\BBP(\ell(f_n(X^n))>nR)=
\BBP(X^n\in C_n^c)$.
Hence,
\begin{align*}
\max\left\{\sup_{\Xps\in\clS}
\BBP\big(X^n\in C_n^c\big),\;\frac{|C_n|}{|A|^n}
\right\}
&\leq
	\max\left\{\sup_{\Xps\in\clS}
	\BBP(\ell(f_n(X^n))>nR),\;\frac{2^{nR+1}}{|A|^n}
	\right\}\\
&\leq
	2\max\left\{\sup_{\Xps\in\clS}
	\BBP(\ell(f_n(X^n))>nR),\;\frac{2^{nR}}{|A|^n}
	\right\}.
\end{align*}
Taking the infimum over all $C_n$ on the left-hand side
and over all pairs of $f_n,R$ on the 
right-hand side
and using 
the definitions
of $n^{\sf fl}(\clS,\epsilon)$ and 
$n^*(\clS,\epsilon)$, gives the required lower bound.
This proves~(\ref{eq:uni1}). The proof of~(\ref{eq:uni2}) is similar.

$(ii)$:
The upper bound in~(\ref{eq:epsilonstarpU})
follows trivially from the fact that prefix-free
codes are a subset of all one-to-one compressors.
For the lower bound, given any compressor
$f_n$ and a rate $R>0$, let 
$C_n=\{x^n:\ell(f_n(x^n))\leq nR\}$ as
before, and assume, without loss of generality,
that $C_n$ is nonempty. Since $|C_n|\leq 2^{\lfloor nR\rfloor+1}-1$,
we can construct a prefix-free compressor 
$f_n^{\sf p}$ that maps all elements $x^n\in C_n$
to the (lexicographically) first $|C_n|$
binary strings of length $\lfloor nR\rfloor$+1,
and maps all the rest of the $x^n$ to binary strings
of length at least $\lfloor nR\rfloor+2$, 
all starting with 
$(\lfloor nR\rfloor+1)$ 1s, and while maintaining the
prefix-free property. Then, for any source $\Xp$,
$\BBP(\ell(f_n(X^n))>nR)
=\BBP(\ell(f^{\sf p}_n(X^n))>nR+1)$, and hence,
$$\inf_{f_n}\sup_{\Xps\in\clS}\BBP\big(\ell(f_n(X^n))>nR\big)
\geq
\inf_{f_n^{\sf p}}\sup_{\Xps\in\clS}
\BBP\big(\ell(f^{\sf p}_n(X^n))>nR+1\big),$$
as required.

$(iii)$: The bounds in~(\ref{eq:uni3}) and~(\ref{eq:uni4})
follow from~(\ref{eq:epsilonstarpU}) in exactly the same
way as~(\ref{eq:nstarvsnp})
and~(\ref{eq:NstarvsNp})
in Theorem~\ref{thm:vltopf} 
follow from the relation~(\ref{eq:epsilonstarp})
from~\cite[Theorem~1]{kontoyiannis-verdu:14}
stated in Section~\ref{s:generalf}.
\qed

\subsection{Memoryless sources}
\label{s:univeralIID}

Next, we obtain bounds on the universal sample
complexity of families of memoryless sources.
In view of Theorem~\ref{thm:universalFL}, like in the case
of a known source, it suffices to establish
sample complexity bounds
for just one of the six fundamental limits.
In this section, we find it convenient to state our
results in terms of $n^{\sf fl}(\clS,\epsilon)$.

Let $\clP$ denote the simplex of all p.m.f.s $P$ on
a fixed finite alphabet $A$ of size $|A|=m$.
For an arbitrary family $\clQ\subset\clP$,
with a slight abuse of notation 
we write $n^{\sf fl}(\clQ,\epsilon)$ for
the universal sample complexity $n^{\sf fl}(\clS,\epsilon)$
of the family $\clS$ of memoryless sources $\Xp$
on $A$ with marginal p.m.f.s $P\in\clQ$.

We begin with the simple observation 
that the sample complexity $n^{\sf fl}(\clQ,\epsilon)$
of an arbitrary family 
$\clQ$ 
is at least as large as
the sample complexity $n^{\sf fl}(\Xp,\epsilon)$
of the ``worst'' source $\Xp$ in that family.
The proof of Proposition~\ref{prop:worst}
is based on 
Le Cam's lemma and 
Proposition~\ref{prop:DTV}.

\begin{proposition}[Basic universal sample complexity lower bound]
\label{prop:worst} \!
Let $U$ be the uniform p.m.f.\ on a finite alphabet $A$ of size
$|A|=m$. Let $\clQ\subset\clP$ be an arbitrary
family of p.m.f.s on $A$,
such that $D_{1/2}(\clQ\|U)=\inf_{Q\in\clQ}D_{1/2}(Q\|U)\in(0,\log m)$.
For any $\epsilon\in(0,1)$, the universal fixed-length
sample complexity of $\clQ$ satisfies,
$$
n^{\sf fl}(\clQ,\epsilon) 
\geq
\frac{\log (1/\epsilon)-2}{D_{1/2}(\clQ\|U)}.
$$
\end{proposition}

{\sc Proof.}
Using the elementary minimax inequality gives
$$\inf_{C_n} \left[\sup_{P \in \clQ}P^n(C_n^c) + \frac{|C_n|}{|A|^n}\right]
=
	\inf_{C_n} \sup_{P \in \clQ}[P^n(C_n^c) + U^n(C_n)]
\geq 
	\sup_{P \in \clQ} \inf_{C_n}[ P^n(C_n^c) + U^n(C_n)].
$$
Then, applying
Le Cam's lemma and 
Proposition~\ref{prop:DTV}, we have:
\begin{align*}
\inf_{C_n} \left[\sup_{P \in \clQ}P^n(C_n^c) + \frac{|C_n|}{|A|^n}\right]
&\geq 
	\sup_{P \in \clQ}[1 - \|P-U\|_{\rm TV}] \\
&\geq 
	\sup_{P \in \clQ} 2^{-nD_{1/2}(P\|U) - 1} \\
&= 
	2^{-nD_{1/2}(\clQ\|U) - 1}.
        \end{align*}
Therefore, 
$$N^{\sf fl}(\clQ,\epsilon)\geq  
\frac{\log(1/\epsilon)-1}{D_{1/2}(\clQ\|U)},$$
and the lower bound in the theorem follows 
the fact that $n^{\sf fl}(\clQ,\epsilon)\geq N^{\sf fl}(\clQ,2\epsilon)$.
\qed

Next we examine particular collections
of families $\clQ$ that are more structured,
allowing for tight upper and lower bounds 
on $n^{\sf fl}(\clQ,\epsilon)$. First, observing
that $n^{\sf fl}(\clQ,\epsilon)$ is exactly
the sample complexity of the identity testing
problem ``$P=U$ vs.\ $P\in\clQ$'',  
\cite[Theorem~2]{diakonikolas:18} immediately
gives the following bounds for the universal
sample complexity of the total variation
families,
$$
\clQ_{{\rm TV},\delta}=
\big\{P\in\clP:\|P-U\|_{\rm TV}\geq \delta\big\},
\quad \delta\in(0,1).
$$

\begin{theorem}
[Universal sample complexity of TV families~\cite{diakonikolas:18}]
\label{thm:diakonikolas}
Let $U$ denote the uniform p.m.f.\ on a finite alphabet $A$ of size
$|A|=m$. For any $\delta\in(0,1)$
and $\epsilon\in(0,1)$, the universal fixed-length
sample complexity of $\clQ_{{\rm TV},\delta}$ satisfies,
$$
n^{\sf fl}(\clQ_{{\rm TV},\delta},\epsilon)
\asymp
\frac{
\log(1/\epsilon)+\sqrt{m\log(1/\epsilon)}
}
{
\delta^2
}.$$
\end{theorem}

Finally, since the sample complexity 
of a known source $P$ was seen to 
naturally be expressed
in terms of its R\'{e}nyi divergence
$D_{1/2}(P\|U)$ from the uniform, 
we examine the sample complexity
of the corresponding $D_{1/2}$ families,
\bqq
\clQ_{\delta}=
\big\{P\in\clP:D_{1/2}(P\|U)\geq \delta\big\},
\label{eq:DhalfQ}
\eqq
for $\delta\in(0,\log m)$.
The proof of Theorem~\ref{thm:canonne},
given in Appendix~\ref{proofcomposite},
uses a collision-based statistic 
and some of its properties
derived in~\cite{canonne:22b}.

\begin{theorem}[Universal sample complexity of $D_{1/2}$ families]
\label{thm:canonne}
Let $U$ denote the uniform p.m.f.\ on a finite alphabet $A$ of size
$|A|=m$. For any $\epsilon\in(0,1)$ and 
all $\delta\in(0,\log e)$,
the universal fixed-length
sample complexity of $\clQ_{\delta}$ satisfies:
\begin{itemize}
\item[$(i)$]
$\displaystyle{
n^{\sf fl}(\clQ_\delta,\epsilon)
\geq
C_1
\frac{
\log(1/\epsilon)+\sqrt{m\log(1/\epsilon)}
}
{
\delta
}}.$
\item[$(ii)$]
$\displaystyle{
n^{\sf fl}(\clQ_{\delta},\epsilon)
\leq C_2
\frac{\sqrt{m}\big[\log(1/\epsilon)+\frac{1}{18}\big]
}
{
\delta
}}.$
\end{itemize}
The constants $C_1,C_2$ are independent of $m,\epsilon$ 
and $\delta$.
\end{theorem}

\noindent
{\bf Remarks.}
\begin{enumerate}
\item
{\em Matching bounds.}
In general, the upper and lower bounds
in~$(i)$ and~$(ii)$ differ by at most
a factor of $\sqrt{m}$. But in several
asymptotic regimes they match. In
particular, in the {\em moderate confidence}
regime when $\epsilon$ is bounded in
a region away from zero (e.g., if
$\epsilon=0.05$ is fixed),  
both the upper and lower bounds
agree up to constants, and
$n^{\sf fl}(\clQ_\delta,\epsilon)\asymp\sqrt{m}/\delta$.
\item
{\em Separation rates.}
The bounds in the theorem lead to separation
rates interpretations analogous to those
in hypothesis testing. For example, the 
lower bound in~$(i)$ implies that, for a 
fixed blocklength~$n$, the largest family
that can be compressed
to within ``error'' $\epsilon$ must be
separated from the uniform by a 
$D_{1/2}$ distance $\delta$ of at least
$$\delta\geq C_1\frac{\log(1/\epsilon)+\sqrt{m\log(1/\epsilon)}}{n}.$$
Or, in the moderate confidence regime
where $\log(1/\epsilon)$ is bounded, 
$(i)$ and~$(ii)$ imply that 
the minimal separation $\delta_{\rm min}$ satisfies
$$\delta_{\rm min}\asymp\frac{\sqrt{m}}{n}.$$
\item
{\em Range of $\delta$.}
Although Theorem~\ref{thm:canonne}
does not hold for all $\delta$ in the 
range $(0,\log m)$ of possible
values of $D_{1/2}(\clQ\|P)$, it does
hold in the interesting and ``difficult''
regime of small $\delta$ when the
sample complexity blows up.
\item
{\em Sublinear compression.}
As noted earlier in Section~\ref{s:IID},
our sample complexity bounds capture the 
fundamental exponential behavior of
the excess-rate probability and 
the rate. 
In the case of universal compression
they also imply that
the minimal blocklength
$n^{\sf fl}(\clQ_\delta,\epsilon)$ 
required to achieve
optimal compression to within error $\epsilon$
and separation~$\delta$, is {\em sublinear}
in the alphabet size $m$. That is, for large $m$,
optimal compression is already achievable
at blocklengths of order $\sqrt{m}$.
This ``sublinear sample complexity'' phenomenon is not
new. It was first observed by Paninski~\cite{paninski:04}
in the related context of entropy estimation. See~\cite{paninski:04}
for a discussion of both the underlying intuition 
and the relevant mathematical reasoning.
\item
{\em $\chi^2$ families.}
A quick look at the proof of Theorem~\ref{thm:canonne} 
shows that the results in both~$(i)$ and~$(ii)$ remain valid 
exactly as stated for
the families $\clQ_{\chi^2,\delta}$ defined
in terms of the $\chi^2$ divergence 
$$\clQ_{\chi^2,\delta}=\Big\{P\in\clP:\chi^2(P\|U)\geq\delta\Big\},$$
in place
of $\clQ_\delta$. The upper bound 
on $n^{\sf fl}(\clQ_{\chi^2,\delta},\epsilon)$ 
is already established in the proof, and the
lower bound follows from the lower bound
in Theorem~\ref{thm:diakonikolas} combined
with the inequality 
$\|P-U\|^2_{\rm TV}\leq\chi^2(P\|U)$,
which comes from combining~(\ref{eq:TV2D12}) and~(\ref{eq:D12c2}).
\item
{\em R\'{e}nyi entropy.}
As in the case of a known source, the key property
of the family $\clQ_\delta$ that determines its
universal sample complexity is the 
distance $D_{1/2}(\clQ_\delta\|U)$ between $\clQ_\delta$ and~$U$
or, equivalently, the maximum R\'{e}nyi entropy
$\sup_{P\in\clQ_\delta}H_{1/2}(P)=\log m -D_{1/2}(\clQ_\delta\|U)$
among all the p.m.f.s in $\clQ$.
\end{enumerate}

\section{Conclusions and future directions}

We introduced a new framework for the study of the 
fundamental performance limits of lossless data compression.
This is done by adopting the
{\em sample complexity} as the central object of study,
defined as the shortest blocklength 
at which both the excess-rate probability and an 
appropriately normalized rate can be controlled 
simultaneously. The sample-complexity point of view leads to genuinely 
non-asymptotic bounds on the fundamental limits, complementing
the classical asymptotic approximations based on the central
limit theorem and on large deviations. For memoryless sources, 
the sample complexity is governed not by the Shannon entropy, 
but by the R\'{e}nyi divergence of order~1/2 between the source
distribution and the uniform distribution.
The same principle persists, in appropriate forms, for Markov 
sources 
and for the problem of universal compression of memoryless
sources.

The other major theme of this work is the exploration 
of the connection between lossless data compression and 
modern tools and techniques used in hypothesis testing. 
The fixed-length formulation is directly related to simple-versus-simple 
testing between the source distribution and the uniform distribution, 
while the universal compression problem is analogous to composite testing 
and, in particular, to identity or uniformity testing. These connections
facilitate the application of non-asymptotic ideas and results 
from statistics and theoretical computer science 
for source coding, and they lead to a new operational interpretation 
of separation rates in terms of universal compression. 

There are several natural directions for further work. 
First, several open problems have already implicitly been mentioned
earlier.
One is to develop sharper bounds 
than those in Theorem~\ref{thm:irreducible} for the sample 
complexity of compressing general ergodic Markov chains.
Another is to further pursue the asymmetric formulations 
introduced in Section~\ref{s:asymmetric}, where the rate constraint 
and 
the excess-rate probability are not required to be controlled 
on the same scale. It would also be interesting to obtain
more closely matching upper and lower bounds for the sample
complexity of universally compressing $D_{1/2}$-families
as in Theorem~\ref{thm:canonne}.

More broadly speaking, we propose that the sample-complexity
formulation can be adopted and studied in connection
with any one of the many classical information-theoretic
problems -- including rate-distortion theory, multiterminal 
source coding, channel coding, and beyond -- to provide 
strong and insightful results on the relevant fundamental
limits. In this vein, we have obtained some encouraging
preliminary results for the problem of efficient communication
over discrete memoryless channels~\cite{viaud-C:manuscript}.

\newpage

\begin{appendices}

\section{Proof of Proposition~\ref{prop:DTV}}
\label{app:DTV}

First, note that, from the definition of $D_{1/2}(P\|Q)$,
\bqq
\sum_{x\in A} \big( \sqrt{P(x)} - \sqrt{Q(x)} \big)^2 
= 2 - 2 \sum_{x\in A} \sqrt{P(x)Q(x)} 
= 2 \left( 1 - 2^{-\frac{1}{2}D_{1/2}(P \|Q)} \right),
\label{eq:sums1}
\eqq
and, similarly,
\bqq
\sum_{x\in A} \big( \sqrt{P(x)} + \sqrt{Q(x)} \big)^2 
= 2 + 2 \sum_{x\in A} \sqrt{P(x)Q(x)} 
= 2 \left( 1 + 2^{-\frac{1}{2}D_{1/2}(P \|Q)} \right).
\label{eq:sums2}
\eqq
We also trivially have, for any $x\in A$: 
\bqq
| P(x) - Q(x) | 
= \big| \sqrt{P(x)} - \sqrt{Q(x)}\big| 
\big( \sqrt{P(x)} + \sqrt{Q(x)} \big) 
\geq \big( \sqrt{P(x)} - \sqrt{Q(x)} \big)^2.
\label{eq:squares}
\eqq
Combining~(\ref{eq:squares}) with~(\ref{eq:sums1}) gives 
the claimed upper bound:
$$1 - \|P-Q\|_{\rm TV} 
= 1 - \frac{1}{2} \sum_{x\in A} 
| P(x) - Q(x) | 
\leq 1 - \frac{1}{2} \sum_{x\in A} \big( \sqrt{P(x)} - \sqrt{Q(x)}\big)^2 
= 2^{-\frac{1}{2}D_{1/2}(P \|Q)}.$$

On the other hand, by the Cauchy-Schwarz inequality
and the identities~(\ref{eq:sums1}) and~(\ref{eq:sums2}),
\begin{align*}
\left[\sum_{x\in A} |P(x) - Q(x) |\right]^2 
&= 
	\left[
	\sum_{x\in A}\big|\sqrt{P(x)}-\sqrt{Q(x)}\big|
	\big(\sqrt{P(x)} + \sqrt{Q(x)} \big)\right]^2 
	\\ 
&\leq 
	\sum_{x\in A} \big( \sqrt{P(x)} - \sqrt{Q(x)} \big)^2 
	\sum_{x\in A} \big( \sqrt{P(x)} + \sqrt{Q(x)} \big)^2 
	\\
&= 4 \Big[ 1 - 2^{-D_{1/2}(P \| Q)}\Big]
\end{align*}
Therefore,
\bqq
1 - \|P-Q\|_{\rm TV}
=1-\frac{1}{2}\sum_{x\in A}|P(x)-Q(x)|
\geq 1 - \Big[1 - 2^{-D_{1/2}(P \| Q)}\Big]^{1/2}
\geq 2^{-D_{1/2}(P \| Q)-1},
\label{eq:DTVapp}
\eqq
where the last inequality follows from the elementary
bound $1 - \sqrt{1-t}\geq \frac{t}{2}$ for $t\geq 0$.
This gives the desired lower bound and completes the proof.
\qed

\section{Proof of Theorem~\ref{thm:canonne}} 
\label{proofcomposite}

The lower bound in~$(i)$ is an
easy consequence of the corresponding
lower bound in Theorem~\ref{thm:diakonikolas}.
Rearranging the inequality~(\ref{eq:DTVapp}) established 
in the proof of Proposition~\ref{prop:DTV} above,
we have,
\bqq
\|P-U\|_{\rm TV}^2 \leq 1 - e^{-(\log_e 2)D_{1/2}(P \| U)} 
\leq (\log_e 2) D_{1/2}(P \| U),
\label{eq:TV2D12}
\eqq
where the last bound follows from the 
elementary inequality $1-e^{-t} \leq t$, $t \in \mathbb{R}$. 
Hence,
for any $\delta\in(0,\log e)$ and $\epsilon\in(0,1)$, we have,
$$
n^{\sf fl}\Big(\clQ_{{\rm TV},\sqrt{\frac{\delta}{\log e}}},\epsilon\Big)
\leq
n^{\sf fl}(\clQ_\delta,\epsilon).
$$
This together with the lower bound in
Theorem~\ref{thm:diakonikolas} prove~$(i)$.

The proof of the upper bound in~$(ii)$
is a consequence of the performance of 
collision-based estimators and a standard
``amplification'' method.
First, recall that
\bqq
D_{1/2}(P\|U)\leq D(P\|U)
\leq \log\big(1+\chi^2(P\|U)\big)\leq(\log e)\chi^2(P\|U).
\label{eq:D12c2}
\eqq
Therefore, to get an upper bound on $n^{\sf fl}(\clQ_\delta,\epsilon)$
it suffices to get an upper bound 
on $n^{\sf fl}\big(\clQ_{\chi^2,\frac{\delta}{\log e}},\epsilon\big)$
where, for any $\zeta$, 
$$\clQ_{\chi^2,\zeta} = \left\{ P\in\clP:\chi^2(P\|U)\geq \zeta \right\}.$$
As $\chi^2(P\|U) = m  \|P\|_2^2 - 1$, the collision-based 
statistic
$$Z_n(x^n) = m \binom{n}{2}^{-1}\sum_{1 \leq i < j \leq n}
\IND{\{x_i = x_j\}}- 1,\quad x^n\in A^n,$$
is an unbiased estimator of $\chi^2(P\|U)$
for any $P\in\clP$. 
For any $\delta\in(0,m-1)$, let
$$C^*_n = \left\{ x^n \in A^n : Z_n(x^n) > \frac{\delta}{2} \right\}.$$
We will obtain an upper bound on $n^{\sf fl}(\clQ_{\chi^2,\delta},\epsilon)$
by bounding $U^n(C_n)$ and $P^n(C_n^{*c})$ for $P\in\clQ_{\chi^2,\delta}$.

Fix a $\delta\in(0,1)$.
Following along the same lines as the proof 
of~\cite[Theorem~2.1]{canonne:22b}, we 
recall from~\cite[Eq.~(2.9)]{canonne:22b}
that, with $X^n\sim P^n$ for any $P$, we have
\bqq
\VAR(Z_n(X^n)) = 
\frac{4m^2}{n^2} \|P\|_2^2 
+ \frac{4m^2}{n} \left( \|P\|_3^3 - \|P\|_2^4 \right).
\label{eq:variance}
\eqq
When $P=U$, we have
$\|U\|_3^3 - \|U\|_2^4 = 0$ and $\|U\|_2^2 = \frac{1}{m}$.
Then Chebyshev's inequality combined with~(\ref{eq:variance}) gives
$$U^n \left( C_n \right) \leq \frac{4\text{Var}(Z_n(X^n))}{\delta^2} 
\leq \frac{16m}{\delta^2n^2}.$$
Therefore, for $n\geq\frac{7\sqrt{m}}{\delta}$, we have
$U^n(C_n)\leq 1/3$.

Now, suppose $X^n\sim P^n$ for some $P$ with
$\alpha=\chi^2(P\|U)\geq\delta$. Using
Chebyshev's inequality and~(\ref{eq:variance}) again,
\begin{align}
    P^n ( C_n^{*c} ) 
&= \BBP \Big( Z_n(X^n) - \alpha \leq \frac{\delta-2\alpha}{2} \Big) 
	\nonumber\\
    &\leq \frac{4}{(\delta-2\alpha)^2} \text{Var}(Z_n(X^n)) 
	\nonumber\\
    &\leq \frac{16m^2}{n(\delta-2\alpha)^2} 
	\Big[
	\frac{1}{n} \|P\|_2^2 + \left( \|P\|_3^3 - \|P\|_2^4 \right) 
	\Big].
	\label{eq:chebychev}
\end{align}
Since $\alpha=\chi^2(P\|U) = m\|P\|_2^2-1$, 
for the first term in~(\ref{eq:chebychev})
we have
\begin{align}
    \frac{16m^2}{n^2(\delta-2\alpha)^2}\|P\|_2^2 
    &= \frac{16m \left(1+\alpha\right)}{n^2(\delta-2\alpha)^2}
	\nonumber\\
    &\leq \frac{16m(1+\delta)}{n^2\delta^2}
	\nonumber\\
    &\leq \frac{32m}{n^2\delta^2}.
	\label{eq:term1}
\end{align}
The first inequality above follows
from the fact that 
$\alpha \mapsto \frac{1+\alpha}{(\delta-2\alpha)^2}$ is 
nonincreasing for $\alpha\geq\delta$,
and the second inequality follows from the
fact that $\delta \leq 1$.
For the second term of the sum in~(\ref{eq:chebychev}), 
from~\cite[Eq.~(2.11)]{canonne:22b} we have
$$\|P\|_3^3 - \|P\|_2^4 
\leq \frac{\alpha^{3/2}}{m^{3/2}} + \frac{3\alpha}{m^2},$$
and hence, since $\alpha\geq \delta$,
\begin{align}
    \frac{16m^2}{n(\delta-2\alpha)^2} \left( \|P\|_3^3 - \|P\|_2^4 \right) 
&\leq \frac{16m^2}{n\alpha^2} \left( \frac{\alpha^{3/2}}{m^{3/2}} 
	+ \frac{3\alpha}{m^2} \right) \nonumber\\
    &\leq \frac{16m^{1/2}}{n\delta^{1/2}} + \frac{48}{n\delta} 
	\nonumber\\
    &\leq \frac{64m^{1/2}}{n\delta}.
	\label{eq:term2}
\end{align}
Combining the bounds~(\ref{eq:term1}) and~(\ref{eq:term2}) we obtain that
$$P^n \left( C_n^{*c} \right) 
\leq \frac{32m}{n^2\delta^2} + \frac{64m^{1/2}}{n\delta},$$
and hence $P^n(C_n^{*c})\leq 1/3$ for $n\geq \frac{200\sqrt{m}}{\delta}$.

Thus we have 
$$\max \left\{ \sup\limits_{P \in \clQ_{\chi^2,\delta}} 
P^n \left( C_n^{*c} \right), U^n \left( C_n \right) \right\} 
\leq \frac{1}{3}\quad\mbox{for}\;n \geq \frac{200\sqrt{m}}{\delta},$$
and replacing $\delta$ by $\delta/(\log e)<1$ as in the
theorem's statement
this implies
$$n^{\sf fl}(\clQ_\delta,1/3)\leq 
n^{\sf fl}\Big(\clQ_{\chi^2,\frac{\delta}{\log e}},1/3\Big)\leq
200(\log e)\times\frac{\sqrt{m}}{\delta}.$$
Then a standard amplification argument 
(see, e.g.,~\cite[Lemma~1.1]{canonne:22b}) 
allows us to conclude that, for all $\epsilon\in(0,1)$, 
$$n^{\sf fl}(\clQ_\delta,\epsilon)\leq 200(\log e)\times
\frac{\sqrt{m}\lceil 18\log(1/\epsilon)\rceil}{\delta},$$
and the upper bound in~$(ii)$ follows.
\qed

\section{Proof of Proposition~\ref{prop:actual}}
\label{app:actual}
Write $m=|A|$ for the alphabet size.
In view of the definition of $R_{\rm act}(\epsilon)$, 
it suffices to
show that:
\be
\lim_{\epsilon\to 0}\frac{\log(1/\epsilon)}{n^*(\Xp,\epsilon)} = \clC(P,U).
\label{eq:target0}
\ee

Recall that, by standard error-exponents
results~\cite{jelinek:book,blahut:74,csiszar:book}
combined with~\cite[Theorem~1]{kontoyiannis-verdu:14},
for each $R\in(H(P),\log m)$ we have
$$\lim_{n\to\infty} -\frac{1}{n}\log\epsilon^*(n,R)= E(R),$$
where $E(R)$ is the continuous function given by
$E(R)=D(P_{\alpha^*}\|P)$ with $\alpha^*\in(0,1)$
chosen such that
$H(P_{\alpha^*})=R$ and with the p.m.f.\ $P_\alpha^*$ defined
as in~\eqref{eq:Palpha}.

For each $R\in(H(P),\log m)$, define
$\phi(R)=\min\{E(R),\log m-R\}$,
and 
$$
\clD(P)=\sup_{R\in(H(P),\log m)}\phi(R).
$$
In order to prove~\eqref{eq:target0},
we will show that
\be
\lim_{\epsilon\to 0}\frac{\log(1/\epsilon)}{n^*(\Xp,\epsilon)}=\clD(P).
\label{eq:target1}
\ee
The result of the proposition then follows upon
recalling that $\clD(P)$ is one of the standard
equivalent expressions for $\clC(P,U)$; see,
e.g.,~\cite[Ch.~11]{cover:book2} or~\cite{vanerven:14}.

We begin with the lower bound in~\eqref{eq:target1}.
Let $\eta>0$ and $R\in(H(P),\log m)$ arbitrary. Then, for  
all sufficiently large~$n$,
\be
\epsilon^*(n,R)\le 2^{-n(E(R)-\eta)}.
\label{eq:LDs}
\ee
Writing
$
n_\epsilon
=
\left\lceil
\frac{\log(1/\epsilon)}{\phi(R)-\eta}
\right\rceil,
$
for all $\epsilon>0$ small enough the bound~\eqref{eq:LDs}
holds with $n_\epsilon$ in place of $n$.
Also, by the definitions of $n_\epsilon$ and $\phi$,
we have that both
$n_\epsilon[E(R)-\eta]\geq\log(1/\epsilon)$
and
$n_\epsilon[\log m-R]\geq\log(1/\epsilon)$.
Hence, for $\epsilon>0$ small enough we have
that both
$$
\epsilon^*(n_\epsilon,R)\leq \epsilon
\quad\mbox{and}\quad
2^{-n_\epsilon[\log m-R]}\leq \epsilon,
$$
which implies that $n^*(\Xp,\epsilon)\le n_\epsilon.$
Therefore,
$$
\liminf_{\epsilon\to 0}
\frac{\log(1/\epsilon)}{n^*(\Xp,\epsilon)}
\geq
\liminf_{\epsilon\to 0}
\frac{\log(1/\epsilon)}{n_\epsilon}
\geq
\phi(R)-\eta,
$$
and since $R\in(H(P),\log m)$ and $\eta>0$ were arbitrary, 
\be
\liminf_{\epsilon\to 0}
\frac{\log(1/\epsilon)}{n^*(\Xp,\epsilon)}
\ge \clD(P).
\label{eq:ALB}
\ee

For the corresponding upper bound, suppose it fails,
that is, 
that there is an $\eta>0$ and a positive sequence 
$\{\epsilon_k\}$ which decreases to zero such that, 
$$\frac{\log(1/\epsilon_k)}{n^*(\Xp,\epsilon_k)}\geq \clD(P)+\eta,
\quad \mbox{for all}\; k.$$
For each $k$, choose a $R_k$ satisfying the constraint
in the definition of $n^*(\Xp,\epsilon_k)$. Then
we have that both
$
\epsilon^*\big(n^*(\Xp,\epsilon_k),R_k\big)\leq \epsilon_k,
$
and
$$
2^{-n^*(\Xps,\epsilon_k)[\log m-R_k]}\leq\epsilon_k,
$$
thus, rearranging,
\be
-\frac{1}{n^*(\Xp,\epsilon_k)}\log\epsilon^*\big(n^*(\Xp,\epsilon_k),R_k\big)
\geq
\frac{\log(1/\epsilon_k)}{n^*(\Xp,\epsilon_k)}
\geq \clD(P)+\eta,
\label{eq:willfail}
\ee
and
$$
\log m-R_k\geq \frac{\log(1/\epsilon_k)}{n^*(\Xp,\epsilon_k)}\geq
\clD(P)+\eta.
$$
In particular,
$ R_k\leq \log m-\clD(P)-\eta$.

Now, since $E(R)$ decreases to zero as $R$ decreases to
$H(P)$, we can find an 
$R_0\in(H(P),\log m)$ such that
$E(R_0)<\clD(P)+\frac{\eta}{2}$.
Next we show that the rates $R_k$
above eventually satisfy $R_k>R_0$.
If $R_k\leq R_0$ along an infinite subsequence
then by the monotonicity of $\epsilon^*(n,R)$ in $R$,
$$
\epsilon^*\big(n^*(\Xp,\epsilon_k),R_k\big)\geq 
\epsilon^*\big(n^*(\Xp,\epsilon_k),R_0\big),
$$
and hence,
$$
-\frac{1}{n^*(\Xp,\epsilon_k)}\log\epsilon^*\big(n^*(\Xp,\epsilon_k),R_k\big)
\le
-\frac{1}{n^*(\Xp,\epsilon_k)}\log\epsilon^*\big(n^*(\Xp,\epsilon_k),R_0\big).
$$
Since $\epsilon_k\to 0$ as $k\to\infty$, 
we have $n^*(\Xp,\epsilon_k)\to\infty$, 
and
$$
\limsup_{k\to\infty}
\Big[
-\frac{1}{n^*(\Xp,\epsilon_k)}\log\epsilon^*\big(n^*(\Xp,\epsilon_k),R_k\big)
\Big]
\leq
E(R_0)
<
\clD(P)+\frac{\eta}{2},
$$
contradicting~\eqref{eq:willfail}.
Therefore, we must eventually have
$R_k\in [R_0,\log m-\clD(P)-\eta]$.

Suppose the sequence $\{R_k\}$ converges to some
$R_*\in [R_0,\log m-\clD(P)-\eta]$,
or pick an appropriate subsequence that does.
Then, by the definitions of $\clD(P)$
and $\phi(R)$ we have
$E(R_*)\leq\clD(P)$.
By the continuity of $E(R)$, we may 
choose a $\delta>0$ sufficiently small so that
$R_*+\delta<\log m$
and also
$E(R_*+\delta)<\clD(P)+\frac{\eta}{2}$.
And by the choice of $\{R_k\}$ we have
$R_k\le R_*+\delta$ for
all large enough $k$.
Since $\epsilon^*(n,R)$ is nonincreasing in $R$ we have,
$$
-\frac{1}{n^*(\Xp,\epsilon_k)}\log\epsilon^*\big(n^*(\Xp,\epsilon_k,)R_k\big)
\leq
-\frac{1}{n^*(\Xp,\epsilon_k)}\log\epsilon^*\big(n^*(\Xp,\epsilon_k),
R_*+\delta\big),
$$
and hence,
$$
\limsup_{k\to\infty}
\Big[
-\frac{1}{n^*(\Xp,\epsilon_k)}\log\epsilon^*\big(n^*(\Xp,\epsilon_k),R_k\big)
\Big]
\leq
E(R_*+\delta)
<
\clD(P)+\frac{\eta}{2},
$$
which contradicts~\eqref{eq:willfail}, as desired.
Therefore, we have,
\be
\limsup_{\epsilon\to 0}
\frac{\log(1/\epsilon)}{n^*(\Xp,\epsilon)}
\le \clD(P),
\label{eq:AUB}
\ee
and combining~\eqref{eq:ALB} with~\eqref{eq:AUB} 
gives~\eqref{eq:target1} as required.
\qed

\end{appendices}

\newpage

\bibliographystyle{plain}
\bibliography{ik}

@ARTICLE{kontoyiannis-verdu:14,
author={Kontoyiannis, I. and Verd\'{u}, S.},
journal =     {IEEE Trans. Inform. Theory},
title={Optimal lossless data compression: {N}on-asymptotics and
asymptotics},
year={2014},
month={February},
volume={60},
number={2},
pages={777-795},
}

@INPROCEEDINGS{kontoyiannis-verdu:13,
  author={Kontoyiannis, I. and Verd\'{u}, S.},
  booktitle={2013 IEEE International Symposium on Information Theory (ISIT)}, 
  title={Optimal lossless compression: Source varentropy and dispersion}, 
  address={Istanbul, Turkey},
  month={July},
  year={2013},
  pages={1739-1743},
}

@article{viaud-C:manuscript,
  title={The sample complexity of channel coding},
  author={Viaud, T. and Kontoyiannis, I.},
  journal = {Manuscript},
  month={June},
  year={2026}
}

@article{theocharous:toappear,
  title={Pragmatic lossless compression:
	{Fundamental} limits and universality},
  author={Theocharous, A. and Gavalakis, L. and Kontoyiannis, I.},
  journal =     "IEEE Trans. Inform. Theory",
  volume = {To appear},
  month={},
  year={2026}
}

@Article{kontoyiannis-97,
  author =      "Kontoyiannis, I.",
  title =       "Second-order noiseless source coding theorems",
  journal =     "IEEE Trans. Inform. Theory",
  year =        "1997",
  month =       "July",
  pages =       "1339-1341",
  volume =      "43",
  number =      "4"
}

@INPROCEEDINGS{altug-wagner-K:13,
  author={Altu{\u{g}}, Y. and Wagner, A.B. and Kontoyiannis, I.},
  title={Lossless compression with moderate error probability}, 
  year={2013},
  pages={1744-1748},
  booktitle={2013 IEEE International Symposium on Information Theory (ISIT)},
  month={July},
  address={Istanbul, Turkey},
}

@book{wald:book,
  title={Statistical decision functions},
  author={Wald, A.},
  year={1950},
  publisher={Wiley},
  address={New York, NY},
}

@book{lecam:86,
  title={Asymptotic methods in statistical decision theory},
  author={Le Cam, L.},
  year={1986},
  publisher={Springer-Verlag},
  address={New York, NY}
}

@book{seneta:81,
  title={Non-negative matrices and {Markov} chains},
  author={Seneta, E.},
  year={1981},
  publisher={Springer},
  address={New York, NY}
}

@Book{csiszar:book,
  Author =      "Csisz{\'{a}}r, I. and K{\"{o}}rner, J.",
  title =       "Information theory: Coding 
		 theorems for discrete memoryless systems",
  publisher =   "Academic Press",
  year =        "1981",
  address =     "New York, NY",
  summary =     ""
}

@Book{cover:book2,
  author =      "Cover, T.M. and Thomas, J.A.",
  title =       "Elements of information theory",
  publisher =   "John Wiley \& Sons",
  year =        "2006",
  address =     "New York, NY",
  edition = 	"Second",
  summary =     ""
}

@Book{jelinek:book,
  title={Probabilistic information theory: {Discrete} and memoryless models},
  author={Jelinek, F.K.},
  year={1968},
  publisher={McGraw-Hill},
  address={New York, NY}
}

@phdthesis{bar-yossef:02,
  title={The complexity of massive data set computations},
  author={Bar-Yossef, Z.},
  year={2002},
  school={Department of Computer Science, University of California, Berkeley},
  address={Berkeley, CA}
}

@inproceedings{lecam:56,
  title={On the asymptotic theory of estimation and testing hypotheses},
  author={Le Cam, L.},
  booktitle={Proc. 3rd Berkeley Sympos. Math. Statist. and Probab.},
  volume={3},
  pages={129-157},
  year={1956},
  organization={University of California Press},
  address={Berkeley, CA},
}

@article{canonne:22,
  author={Canonne, C.L.},
  title={A short note on an inequality between {KL} and {TV}},
  journal = {arXiv e-prints},
  volume = {\texttt{2202.07198 [math.PR]}},
  month={February},
  year={2022}
}

@book{canonne:20,
 author = {Canonne, C.L.},
  title={A survey on distribution testing: {Your} data is big. 
	{But} is it blue?},
 year = {2020},
 pages = {1-100},
 publisher = {Theory of Computing Library},
 number = {9},
 series = {Graduate Surveys},
}

@article{canonne:22b,
  title={Topics and techniques in distribution testing: 
	{A} biased but representative sample},
  author={Canonne, C.L.},
  journal={Foundations and Trends in Communications and Information Theory},
  publisher={Now Publishers},
  volume={19},
  number={6},
  year={2022},
  pages={1032-1198},
  month={November}
}

@article{csiszar-shields:04,
    AUTHOR = {Csisz{\'{a}}r, I. and Shields, P.},
     TITLE = {Information theory and statistics: {A} tutorial},
   journal = {Foundations and Trends in Communications and Information Theory},
   publisher={NOW Publishers},
    VOLUME = {1},
     number = {4},
      YEAR = {2004},
   pages={417-528},
     MONTH = {December},
}

@article{ingster:84,
  title={Asymptotically minimax testing of nonparametric hypotheses on 
	the density of the distribution of an independent sample},
  author={Ingster, Yu.I.},
  journal={Zap. Nauchn. Sem. Leningrad. Otdel. Mat. Inst. Steklov.(LOMI)},
  volume={136},
  pages={74-96},
  year={1984}
}

@article{ingster:85,
  title={Asymptotically optimal tests for composite finite-parametric 
	hypotheses},
  author={Ingster, Yu.I.},
  journal={Teoriya Veroyatnostei i ee Primeneniya},
  volume={30},
  number={2},
  pages={289-308},
  year={1985},
}

@book{ingster:book,
  title={Nonparametric goodness-of-fit testing under 
	{Gaussian} models},
  author={Ingster, Yu.I. and Suslina, I.A.},
  year={2002},
  series={Lecture Notes in Statistics},
  volume={169},
  publisher={Springer},
  address={New York, NY},
}

@article{lecam:60b,
  title={Locally asymptotically normal families of distributions. 
	{Certain} approximations to families of distributions and their 
	use in the theory of estimation and testing hypotheses},
  author={Le Cam, L.},
  journal={Univ. California Publ. Statist., {\em Berkeley, CA}},
  volume={3},
  pages={37},
  year={1960}
}

@INPROCEEDINGS{valiant:14,
  author={Valiant, G. and Valiant, P.},
  booktitle={2014 IEEE 55th Annual Symposium 
	on Foundations of Computer Science}, 
  title={An Automatic Inequality Prover and Instance Optimal Identity Testing}, 
  year={2014},
  volume={},
  number={},
  pages={51-60},
	}

@inproceedings{diakonikolas:15,
author = {Diakonikolas, I. and Kane, D.M. and Nikishkin, V.},
title = {Testing Identity of Structured Distributions},
booktitle = {Proceedings of the 2015 Annual ACM-SIAM Symposium 
	on Discrete Algorithms (SODA)},
chapter = {},
pages = {1841-1854},
publisher={SIAM},
year={2015},
}

@InProceedings{diakonikolas:18,
  author =	{Diakonikolas, I. and Gouleakis, T. and Peebles, J.
			 and Price, E.},
  title =	{Sample-Optimal Identity Testing with High Probability},
  booktitle =	{45th International Colloquium on Automata, Languages, 
		and Programming (ICALP 2018)},
  pages =	{41:1-14},
  series =	{Leibniz International Proceedings in Informatics (LIPIcs)},
  year =	{2018},
  volume =	{107},
  editor =	{Chatzigiannakis, I. and Kaklamanis, C. and Marx, D.
			and Sannella, D.},
  address =	{Dagstuhl, Germany},
}

@article{diakonikolas:19,
  title={Collision-based testers are optimal for uniformity and closeness},
  author={Diakonikolas, I. and Gouleakis, T. and Peebles, J. and Price, E.},
  journal={Chic. J. Theor. Comput. Sci},
  volume={25},
  pages={1-21},
  year={2019}
}

@InProceedings{diakonikolas:19b,
  title = 	 {Communication and Memory Efficient Testing 
		of Discrete Distributions},
  author =       {Diakonikolas, I. and Gouleakis, T. and Kane, D.M. and 
		Rao, S.},
  booktitle = 	 {32nd Conference on Learning Theory (COLT)},
  pages = 	 {1070-1106},
  year = 	 {2019},
  editor = 	 {Beygelzimer, A. and Hsu, D.},
  volume = 	 {99},
  series = 	 {Proceedings of Machine Learning Research},
  month = 	 {June},
}

@Incollection{goldreich:11,
	author="Goldreich, O. and Ron, D.",
	editor="Goldreich, O.",
	title="On Testing Expansion in Bounded-Degree Graphs",
	bookTitle="Studies in Complexity and Cryptography. Miscellanea on the 
	Interplay between Randomness and Computation",
year="2011",
publisher="Springer",
address="Berlin, Heidelberg",
pages="68-75",
}

@inproceedings{acharya:15,
 title = {Optimal Testing for Properties of Distributions},
 author = {Acharya, J. and Daskalakis, C. and Kamath, G.},
 booktitle = {Advances in Neural Information Processing Systems},
 volume = {28},
 editor = {C. Cortes and N. Lawrence and D. Lee and M. Sugiyama and R. Garnett},
 year = {2015},
  month={December},
  address={Montr\'{e}al, Quebec},
}

@ARTICLE{acharya:20,
  author={Acharya, J. and Canonne, C.L. and Tyagi, H.},
   JOURNAL = {IEEE Trans. Inform. Theory},
  title={Inference Under Information Constraints {II: Communication}
	 Constraints and Shared Randomness}, 
  year={2020},
  volume={66},
  number={12},
  pages={7856-7877},
  month={October},
}

@INPROCEEDINGS{acharya:21,
  author={Acharya, J. and Canonne, C.L. and Liu, Y. and Sun, Z.
	and Tyagi, H.},
  booktitle = "2021 IEEE International Symposium on Information Theory (ISIT)",
  title={Interactive Inference under Information Constraints}, 
  year={2021},
  pages={326-331},
  address={Melbourne, Australia},
  month={July},
}

@ARTICLE{paninski:08,
  author={Paninski, L.},
   JOURNAL = {IEEE Trans. Inform. Theory},
  title={A Coincidence-Based Test for Uniformity Given Very Sparsely 
	Sampled Discrete Data}, 
  year={2008},
  volume={54},
  number={10},
  pages={4750-4755},
  month = {October},
}

@article {blahut:74,
   AUTHOR = {Blahut, R.E.},
    TITLE = {Hypothesis testing and information theory},
  JOURNAL = {IEEE Trans. Inform. Theory},
   VOLUME = {20},
   NUMBER = {4},
     YEAR = {1974},
    PAGES = {405-417},
    MONTH = {July},
}

@article {paninski:04,
    AUTHOR = {Paninski, L.},
     TITLE = {Estimating entropy on {$m$} bins given fewer than {$m$}
              samples},
   JOURNAL = {IEEE Trans. Inform. Theory},
    VOLUME = {50},
      YEAR = {2004},
    NUMBER = {9},
     PAGES = {2200-2203},
     MONTH = {August},
}

@preamble{
   "\def\cprime{$'$} "
}

@article{wald:45,
  title={Statistical decision functions which minimize the maximum risk},
  author={Wald, A.},
  journal={Ann. of Math.},
  volume={46},
  number={2},
  pages={265-280},
  year={1945},
  month={April},
}

@article{neyman:33,
  title={On the problem of the most efficient tests of statistical hypotheses},
  author={Neyman, J. and Pearson, E.S.},
  journal={Philos. Trans. Roy. Soc. London A},
  volume={231},
  number={694-706},
  pages={289-337},
  year={1933},
}

@article{vanerven:14,
  title={R{\'e}nyi divergence and {Kullback-Leibler} divergence},
  author={van Erven, T. and Harremo{\"e}s, P.},
  journal =     "IEEE Trans. Inform. Theory",
  volume={60},
  number={7},
  pages={3797-3820},
  year={2014},
  month={July}
}

@Article{mcmillan,
  author =      "McMillan, B.",
  title =       "The basic theorems of Information Theory",
  journal =     "Ann. Math. Statist.",
  year =        "1953",
  volume =      "24",
  number =      "2",
  pages =       "196-219",
  month =       "June",
  OPTnote =        ""
}

@article{wang:23,
  title={Information divergences of {Markov} chains and their applications},
  author={Wang, Y. and Choi, M.C.H.},
  journal = {arXiv e-prints},
  year={2023},
  month={December},
	volume = {\texttt{2312.04863 [cs.IT]}},
}

@ARTICLE{rached:01,
  author={Rached, Z. and Alajaji, F. and Lorne Campbell, L.},
journal={IEEE Trans. Inform. Theory},
  title={{R\'{e}nyi's} divergence and entropy rates for 
	finite alphabet {Markov} sources}, 
  year={2001},
  volume={47},
  number={4},
  pages={1553-1561},
  month={May}
}

@article{gao:26,
   author = {Gao, J. and Chen, S. and Wu, Y. and Liu, L. and Caire, G.
		and Poor, H.V. and Zhang, W.},
    title = "Finite-blocklength information theory",
  journal = {Fundam. Res.}, 
	volume = {6},
	issue = {4},
     year = "2026",
    month = "July",
}

@inproceedings{pensia:24,
  title={The sample complexity of simple binary hypothesis testing},
  author={Pensia, A. and Jog, V. and Loh, P.-L.},
  booktitle={37th Annual Conference on Learning Theory},
  series =  {Proceedings of Machine Learning Research},  
  pages={4205-4206},
  volume={247},
  year={2024},
  note={Full version at {\em arXiv e-prints} \texttt{2403.16981 [math.ST]}} 
}

@InProceedings{daskalakis:18,
  title = 	 {Testing Symmetric {Markov} Chains From a Single Trajectory},
  author =       {Daskalakis, C. and Dikkala, N. and Gravin, N.},
  booktitle = 	 {31st Conference On Learning Theory},
  pages = 	 {385-409},
  year = 	 {2018},
  editor = 	 {Bubeck, S. and Perchet, V. and Rigollet, P.},
  volume = 	 {75},
  series = 	 {Proceedings of Machine Learning Research},
  month = 	 {July},
}

@InProceedings{cherapanamjeri:19,
  title = 	 {Testing Symmetric {Markov} Chains Without Hitting},
  author =       {Cherapanamjeri, Y. and Bartlett, P.L.},
  booktitle = 	 {32nd Conference on Learning Theory},
  pages = 	 {758-785},
  year = 	 {2019},
  editor = 	 {Beygelzimer, A. and Hsu, D.},
  volume = 	 {99},
  series = 	 {Proceedings of Machine Learning Research},
  month = 	 {June},
}

@InProceedings{chan:21,
  title = 	 {Learning and Testing Irreducible {M}arkov Chains via 
	the {$k$-cover} Time},
  author =       {Chan, S.O. and Ding, Q. and Li, S.H.},
  booktitle = 	 {32nd International Conference on Algorithmic Learning Theory},
  pages = 	 {458-480},
  year = 	 {2021},
  editor = 	 {Feldman, V. and Ligett, K. and Sabato, S.},
  volume = 	 {132},
  series = 	 {Proceedings of Machine Learning Research},
  month = 	 {March},
}

@InProceedings{wolfer:20,
  title = 	 {Minimax Testing of Identity to a Reference Ergodic 
	{Markov} Chain},
  author =       {Wolfer, G. and Kontorovich, A.},
  booktitle = 	 {23rd International Conference on Artificial Intelligence 
	and Statistics},
  pages = 	 {191-201},
  year = 	 {2020},
  editor = 	 {Chiappa, S. and Calandra, R.},
  volume = 	 {108},
  series = 	 {Proceedings of Machine Learning Research},
  month = 	 {August},
}

@article{wolfer:21,
author = {Wolfer, G. and Kontorovich, A.},
year = {2021},
month = {February},
pages = {532-553},
title = {Statistical estimation of ergodic {Markov} chain kernel 
	over discrete state space},
volume = {27},
journal = {Bernoulli},
number={1},
}

@article{ziv:97,
  author = {Ziv, J.},
  title = {Back from infinity: {A} constrained resources approach
	to information theory {(Shannon Lecture)}},
  journal={IEEE Information Theory Society Newsletter},
  volume = {48},
  number = {1},
  pages = {21-30},
  year = {1998},
}

@InCollection{strassen:64b,
  author  = "Strassen, V.",
  TITLE   = {Asymptotische {A}bsch\"atzungen in {S}hannons
             {I}nformationstheorie},
BOOKTITLE = {3rd Prague Conf. Information Theory, Statist. Decision
             Functions, Random Processes (Liblice, 1962)},
    PAGES = {689-723},
PUBLISHER = {Publ. House Czech. Acad. Sci.},
  ADDRESS = {Prague},
     YEAR = {1964},
}

@article{dobrushin:62,
  title={Asymptotic bounds of the probability of error for the transmission 
	of messages over a discrete memoryless channel with a symmetric 
	transition probability matrix},
  author={Dobrushin, R.L.},
  journal={Teor. Veroyatnost. i Primenen},
  volume={7},
  pages={283-311},
  year={1962}
}

@article{csiszar:71,
  title={On the error exponent for source coding and for testing simple 
		statistical hypotheses},
  author={Csisz{\'a}r, I. and Longo, G.},
  journal={Studia Sci. Math. Hungar.},
  volume={6},
  pages={181-191},
  year={1971}
}

@Article{shannon:48,
	author= "Shannon, C.E.",
	title=  "A mathematical theory of communication",
	journal="Bell System Tech. J.",
	volume= "27",
   	number= "3",
	pages=  "379-423, 623-656",
	year=   "1948",
	OPTsummary=""
	}

@Article{yushkevich,
  author =      "Yushkevich, A.A.",
  title =       "On limit theorems connected with the concept of the
		 entropy of {M}arkov chains",
  journal =     "Uspehi Mat. Nauk",
  year =        "1953",
  volume =      "8",
  OPTnumber =   "",
  pages =       "177-180",
  month =       "",
  note = 	"(Russian)"
}

\end{document}